\documentclass[runningheads]{llncs}
\usepackage[T1]{fontenc}
\usepackage{graphicx}
\usepackage{amsmath,hyperref}

\usepackage{amsfonts}
\usepackage{array}
\usepackage[caption=false,font=normalsize,labelfont=sf,textfont=sf]{subfig}
\usepackage{textcomp}
\usepackage{xspace}
\usepackage{stfloats}
\usepackage{url}
\usepackage{verbatim}
\usepackage{cite}
\usepackage{times}  
\usepackage{helvet} 
\usepackage{courier}
\usepackage{color}

\usepackage{amssymb}
\usepackage{booktabs}
\usepackage[noend,ruled,linesnumbered]{algorithm2e}
\usepackage{newfloat}
\usepackage{listings}
\usepackage{xcolor}

\def\premethod{FedFair\xspace}
\def\method{FedFDP\xspace}

\newtheorem{assumption}{Assumption}

\SetKwProg{Retract}{}{}{}

\begin{document}

\title{\method: Fairness-Aware Federated Learning with Differential Privacy}


\author{Xinpeng Ling\inst{\star,1}\orcidID{0009-0006-1605-7054} \and
Jie Fu\inst{\star,2}\orcidID{0009-0009-9337-1452}\protect\footnote[0]{$\star$: Those authors contribute equally.} \and
\\Kuncan Wang\inst{3}\orcidID{0009-0004-4756-5938} \and
Huifa Li\inst{1}\orcidID{0009-0009-5806-385X} \and
\\Tong Cheng\inst{1}\orcidID{0009-0001-3120-6103} \and
Zhili Chen\inst{\dagger,1}\orcidID{0000-0002-2231-3652}
\protect\footnote[0]{$\dagger$: Corresponding author.}
}
\authorrunning{Xinpeng et al.}
%

\institute{East China Normal University, Shanghai, China\\
\email{$\{$xpling,huifali,tcheng$\}$@stu.ecnu.edu.cn,zhlchen@sei.ecnu.edu.cn}
\and
Stevens Institute of Technology, Hoboken NJ 07030, USA\\
\email{jfu13@stevens.edu}
\and
Nanyang Technological University, Singapore\\
\email{kuncan001@e.ntu.edu.sg}
}

\maketitle 

\begin{abstract}
Federated learning (FL) is an emerging machine learning paradigm designed to address the challenge of data silos, attracting considerable attention. However, FL encounters persistent issues related to fairness and data privacy. To tackle these challenges simultaneously, we propose a fairness-aware federated learning algorithm called \premethod. Building on \premethod, we introduce differential privacy to create the \method algorithm, which addresses trade-offs among fairness, privacy protection, and model performance. In \method, we developed a fairness-aware gradient clipping technique to explore the relationship between fairness and differential privacy. Through convergence analysis, we identified the optimal fairness adjustment parameters to achieve both maximum model performance and fairness. Additionally, we present an adaptive clipping method for uploaded loss values to reduce privacy budget consumption. Extensive experimental results show that \method significantly surpasses state-of-the-art solutions in both model performance and fairness.
\end{abstract}

\section{Introduction}\label{sec:introduction}

Federated Learning (FL) \cite{mcmahan2017communication} is a distributed machine learning framework that allows clients to collaboratively train a shared model without exposing their respective datasets. After training their local models, each client only transmits model parameters to the central server, rather than the original training data. The server aggregates these parameters to update the global model, which is then sent back to the clients for further training. Due to its ability to improve model performance through collaboration without the need to upload original data, overcoming data silos, FL has received significant attention in recent years \cite{zhu2021fedgen,lee2022fedntd,chen2024multicenter}. Particularly, many studies have achieved significant breakthroughs in terms of model performance \cite{WangJ2020FedNova,li2020fedprox,li2021ditto} and communication costs \cite{wang2019adaptive,ling2024ali} in FL.


To encourage more clients to participate in federated learning, establishing a fairness-aware federated learning algorithm is necessary \cite{li2021survey}. Current fairness-aware machine learning algorithms, such as \cite{dwork2012fairness,cotter2019optimization}, have not fully addressed the issue of client performance disparities in federated learning. Due to non-independent and identically distributed (Non-IID) data, inconsistent client objectives (orange squares) lead to significant performance disparities of the collaboratively trained global model (red square) across different datasets held by various clients. An increasing number of scholars are beginning to focus on this phenomenon of unfair performance disparity \cite{TianLi2020qFFL,li2021ditto,zhao2022dynamic}.

On the other hand, recent studies have pointed out that training parameters may leak data privacy, such as recovering original data \cite{zhu2019deep} or member inference attacks \cite{song2017machine}. Even if only models or gradients are uploaded instead of original data, data leakage can still occur \cite{LucaMelis2022ExploitingUF}. To enhance data privacy in FL, differential privacy (DP) \cite{dwork2014algorithmic} has become the preferred method for protecting privacy in federated learning due to its moderate computational overhead and solid mathematical foundation. Incorporating appropriate DP during the training phase can effectively prevent the accidental leakage of sensitive training information \cite{fu2024differentially}.

Although some works have explored balanced performance fairness and differential privacy in federated learning separately, no prior research has considered them in a unified framework. For the first time, we propose a fairness-aware federated learning with differential privacy. However, balancing differential privacy protection, fairness, and model performance presents a significant challenge. Specifically, there are two main challenges: First, the clipping and noise addition processes in differential privacy have the potential to affect both fairness and model utility in federated learning. Moreover, achieving fairness in federated learning requires uploading loss information to the server, which entails additional privacy budget consumption.

In this paper, the relationship between the fairness loss function and differential privacy in federated learning is analyzed to address the first challenge. We design an fairness-aware gradient clipping
strategy to balance fairness and differential privacy. This gradient clipping strategy not only meets the requirements for fairness but also satisfies the sensitivity control needed for differential privacy. Furthermore, through convergence analysis, we demonstrate that an optimal adjustment parameter can be found to achieve the best model performance and fairness. To address the second challenge, we design an adaptive clipping method for loss uploading, significantly reducing the privacy budget consumption in this part. 

Our main contributions are as follows:

\begin{enumerate}
    \item First, a fairness-aware federated algorithm, \premethod, is proposed. It integrates a novel loss function specifically designed to simultaneously optimize fairness and model performance.
    \item Based on \premethod, we propose an algorithm called \method to further equip the system  with differential privacy. In particular, we designed a fairness-aware gradient clipping strategy that ensures differential privacy and allows for adjustment of fairness in federated learning. In addition, we propose an adaptive clipping method for the additional loss values uploaded by each client to achieve optimal utility.
    \item Furthermore, we conducted a convergence analysis of \method and identified the fairness parameter $\lambda^*$ that results in the fastest convergence. We analyzed the privacy loss of the \method algorithm using the concept of Rényi Differential Privacy (RDP) to verify its compliance with differential privacy requirements.
    \item Finally, through comprehensive empirical evaluations on three public datasets, we confirmed that \premethod and \method are superior to existing solutions in terms of model performance and fairness. What's more, we investigated the impact of clipping bound and noise multiplier on fairness.
\end{enumerate}

In Section~\ref{sec:problem_formulation}, we formalize our problem and introduce the three objectives of our method.
 Section~\ref{sec:effect_fair} presents a  fair federated learning approach—\premethod.
 In Section~\ref{sec:method}, we enhance \premethod by incorporating the DP guarantee, thereby arriving at the proposed \method. A \textit{fairness parameter} $\lambda$ is embedded within \method, and in Section~\ref{sec:guarantees}, we derive the analytical solution for the optimal $\lambda^*$ through convergence analysis and differential methods.
 The privacy analysis of \method are examined in detail in Section~\ref{sec:privacy_analysis}.
 In Section~\ref{sec:experiments}, we conduct extensive experiments on three datasets, comparing six baseline algorithms to rigorously validate the effectiveness of both \premethod and \method.
 Following the review of related work in Section~\ref{sec:related_work}, Section~\ref{sec:conclusion} provides a summary of our contributions.
 Due to space limitations, the detailed procedures of the convergence analysis are included in Appendix~\ref{sec:appendix}. 
\section{Problem Formulation}\label{sec:problem_formulation}
In this work, we propose \method which aims to serve threefold goals as follows. Then, we will elaborate on and formalize how fairness and DP are defined in FL.

\begin{itemize}
  \item {\bf Goal 1 (Fairness)}: Strive to achieve federated learning with the highest possible fairness, where lower values of the federated fairness metric (Equation \ref{equ:fairness}) indicate better fairness.

 \item {\bf Goal 2 (Privacy)}: Implement federated learning under formal differential privacy guarantees (Equation \ref{def:DP}).

 \item {\bf Goal 3 (Utility)}: Ensure that the federated learning model retains strong predictive performance, particularly in terms of classification accuracy.

\end{itemize}
\subsection{Fairness in FL}
We use $\mathbf{w}^i_t$ signifies the model acquired from client $i$ at iteration $t$, while $\mathbf{w}_t$ denotes the server-aggregated model. The objective of FL is to minimize $F(\mathbf{\cdot})$, in other words, to seek:

\begin{align}
\mathbf{w}^* &= \arg \min F(\mathbf{w}), \text{where } F(\mathbf{w}) = \sum_{i=1}^N p_i F_i(\mathbf{w}).\label{eq:F_w}
\end{align}

Here, $F_i(\cdot)$ represents the local loss function of client $i$. The weights are defined as $p_i=\frac{|D_i|}{\sum_j |D_j|}$, $D_i$ represents the dataset of client $i$.

We have defined balanced performance fairness in FL as follows.

\begin{definition}\label{def:balanced_performance_fairness}
(\textbf{Balanced performance fairness \cite{TianLi2020qFFL}})
For trained models $\mathbf{w}_1$ and $\mathbf{w}_2$, we informally recognize that model $\mathbf{w}_1$ provides a more fair solution to the federated learning objective in Equation (\ref{eq:F_w}) than model $\mathbf{w}_2$ if the performance of model $\mathbf{w}_1$ on the $N$ devices is more uniform than the \textbf{performance} of model $\mathbf{w}_2$ on the $N$ devices.
\end{definition}

Differing from [28] that uses the variance of accuracy for "performance", we employ the variance of loss. This enables the metric to be incorporated directly into the objective function, facilitating more straightforward iterative optimization. Equation (\ref{equ:fairness}) represents the specific form of fairness in federated learning, which is the \textbf{weighted variance} of the loss values of global trained model $\mathbf{w}$ across all clients. Essentially, the smaller the value of $\Psi$, the fairer the $\mathbf{w}$ for all clients:

\begin{equation} \label{equ:fairness}
\Psi(\mathbf{w}) =  \sum_{i=1}^N p_i \left(F_i(\mathbf{w})-F(\mathbf{w})\right)^2.
\end{equation}



\subsection{DP guarantee in FL}\label{subsec:dp}
DP is a formal mathematical model that quantifies privacy leakage in data analysis algorithms. It stipulates that alterations to a single record within the training data should not induce significant statistical variations in the algorithm's results.

\begin{definition}
\label{Differential Privacy}
({\bf DP~\cite{dwork2014algorithmic}}). $(\epsilon, \delta)$-DP is achieved by a randomized mechanism $\mathcal{M}: \mathcal{X}^n \rightarrow \mathbb{R}^d$, for any two neighboring databases $D_i, D_i^{\prime} \in \mathcal{X}^n$ that \textbf{differ in only a single data sample}, and $\forall$ $\mathcal{S} \subseteq \text{Range}(\mathcal{R})$:
\end{definition}

\begin{equation}\label{def:DP}
\operatorname{Pr}\left[\mathcal{M}\left(D_i\right) \in \mathcal{S}\right] \leq e^\epsilon \operatorname{Pr}\left[\mathcal{M}\left(D_i^{\prime}\right) \in \mathcal{S}\right]+\delta.
\end{equation}

By adding random noise, we can achieve differential privacy for a function $f: \mathcal{X}^n \rightarrow \mathbb{R}^d$ according to Definition~\ref{Differential Privacy}. The $l_2$-sensitivity determines how much noise is needed and is defined as follow.
 

\textbf{DPSGD.} In gradient-based methods, DPSGD\cite{Abadi2016DeepLearningWithDP} is widely used in privacy preservation. It performs per-sample gradient clipping and noise addition:
\begin{itemize}
    \item Firstly, computing the gradient for data sample $\xi_j$, donated by
    $\mathbf{g}^{i,j}_t = \nabla F_i(\mathbf{w}_t^i,\xi_j)$.
    \item Secondly, clipping the gradient of per-sample to get the sensitivity: 
    \begin{equation}\label{eq:dp_step2}
        \mathbf{\hat{g}}^{i,j}_t = \mathbf{g}^{i,j}_t / \max\left( 1, \frac{\|\mathbf{g}^{i,j}_t\|}{C} \right),
    \end{equation}
    where $C$ represents the clipping bound.
    \item Thirdly, adding Gaussian noise with mean 0 and standard deviation $C \cdot \sigma$ to the sum of clipped gradients, where $\sigma$ is the noise multiplier, $\mathbf{I}$ is the identity matrix: 
    \begin{equation}\label{eq:dp_step3}
        \mathbf{\tilde{g}}^{i}_t = \frac{1}{|\mathcal{B}_i|} \left( \sum_{j=1}^{|\mathcal{B}_i|} \mathbf{\hat{g}}^{i,j}_t + C \sigma \cdot \mathcal{N}(0,\mathbf{I})\right).
    \end{equation}
\end{itemize}
The final step is to proceed with gradient descent as usual. When each client in FL utilizes DPSGD as the optimization algorithm to safeguard their private data, the framework transitions from FL to DPFL.


\section{\premethod: Equipping FL with Fairness}\label{sec:effect_fair}

\begin{algorithm}[!t]
\caption{\premethod}\label{alg:method_without_DP}
\KwIn{loss function $F(\mathbf{w})$. Parameters: learning rate $\eta$, fairness hyperparameter $\lambda$, batch sample ratio $q$, local dataset $D_i$.}
\KwOut{the final trained model $\mathbf{w}_T$}
Initialize $\mathbf{w}_0,F(\mathbf{w}_{0})=\text{Initial()}$\;
\Retract{\textbf{SERVER}}{\label{alg0:line_server}
\For{$t = 0,1,\cdots,T-1$ }
{\label{alg0:line_server_for1}
\For{$i = 1,2,\cdots,N$ parallel }{\label{alg0:line_server_for2}
$\mathbf{w}_{t+1}^i, F_i(\mathbf{w}_{t+1}^i)=\text{LOCAL}\left( \mathbf{w}_{t},F(\mathbf{w}_{t})\right)$\label{alg0:line_server_local_update}\;
}
$\mathbf{w}_{t+1} = \sum_{i=1}^N p_i \mathbf{w}_{t+1}^i$\label{alg0:line_server_modelAggre}\;
$F(\mathbf{w}_{t+1}) = \sum_{i=1}^N p_i F_i(\mathbf{w}_{t+1}^i)$\label{alg0:line_server_lossAggre}\;
}
\Return $\mathbf{w}_T$\label{alg0:line_server_return}\;
}
\Retract{\textbf{LOCAL}}{\label{alg0:line_local}
$\text{Download } \mathbf{w}_t, F(\mathbf{w}_t) \text{ and } \mathbf{w}_t^{i} \leftarrow \mathbf{w}_t$\label{alg0:line_local_download}\;

$\{\mathcal{B}_i\}\leftarrow$ Split $D_i$ to batches\label{alg0:line_local_sample}\;
\For{batch $b\in \{\mathcal{B}_i\}$}{\label{alg0:line_local_for4}
$\Delta_{i}= F_i(\mathbf{w}_t^i,b)-F(\mathbf{w}_t)$\label{alg0:line_local_delta}\;
$\eta_i=\eta\cdot\left(1+\lambda\cdot\Delta_{i}\right)$\label{alg0:line_local_etai}\;
$\mathbf{w}_{t}^{i}=\mathbf{w}_t^{i} - \eta_{i} \cdot \nabla F_i(\mathbf{w}_t^i,b)$\label{alg0:line_local_update}\;
}
$\mathbf{w}_{t+1}^i\leftarrow \mathbf{w}_{t}^i$ , compute $F_i(\mathbf{w}_{t+1}^i)$\;\label{alg0:line_compute_Fi}
\Return $ \mathbf{w}_{t+1}^i, F_i(\mathbf{w}_{t+1}^i)$\label{alg0:line_local_return}\;
}
\end{algorithm}




We previously defined the performance objective in Equation (\ref{eq:F_w}) and balanced performance fairness in Definition \ref{def:balanced_performance_fairness}. Here, we use these to develop a new objective function that informs our proposed fairness-aware federated learning algorithm.

Upon incorporating Equation (\ref{equ:fairness}) into Equation (\ref{eq:F_w}), we introduce a comprehensive objective function that integrates considerations of fairness:

\begin{equation}
    \min_{\mathbf{w}} H(\mathbf{w}) = F(\mathbf{w}) + \frac{\lambda}{2} \sum_{i=1}^{N} p_i \left( F_i(\mathbf{w}) - F(\mathbf{w}) \right)^2.
\end{equation}

For client $i$ during round $t$, the refined objective is articulated as follows:

\begin{equation}\label{eq:fair_obj}
    \min_{\mathbf{w}^i_t} H_i(\mathbf{w}^i_t) = F_i(\mathbf{w}^i_t) + \frac{\lambda}{2} \left( F_i(\mathbf{w}^i_t) - F(\mathbf{w}^i_t) \right)^2,
\end{equation}

where $F(\mathbf{w}^i_t) = \sum_{i=1}^{N} p_i F_i(\mathbf{w}^i_t)$ and $F_i(\mathbf{w}^i_t)$ is determined using the training dataset\protect\footnotemark \footnotetext{We use $F(\mathbf{w}_t)= \sum_{i=1}^N p_i F_i(\mathbf{w}_t^i)$, the reason for $F(\mathbf{w}_t) \neq F(\mathbf{w}_t,\xi_{test})$ is that the test data $\xi_{test}$ from the server cannot be used for training.} from round $t-1$, as delineated in line \ref{alg0:line_compute_Fi} of Algorithm \ref{alg:method_without_DP}, and the $\lambda$ ($\lambda\geq0$) is a hyperparameter used to adjust the degree of fairness~\footnote{When $\lambda=0$, the algorithm becomes traditional federated learning. When $\lambda\rightarrow\infty$, the algorithm only focuses on fairness. However, if $\lambda$ is too large, it will cause very steep gradients which might prevent the model from converging. Fortunately, we theoretically found an optimal value for $\lambda$ in Section \ref{subsec:optimal_lambda}. In addition, as shown in Figure \ref{fig:optimal_lambda}, we experimentally confirmed the existence of the optimal $\lambda$.}. Consequently, within the context of Equation (\ref{eq:fair_obj}), $F(\mathbf{w}^i_t)$ is considered a constant when the gradient is being computed. So, the gradient of Equation (\ref{eq:fair_obj}) as follows:
\begin{align}\label{eq:grad_Hi_batch}
    \nabla H_i(\mathbf{w}_t^i)&=\nabla F_i(\mathbf{w}_t^i) + \lambda\Delta_{i}\cdot\nabla F_i(\mathbf{w}_t^i)
    \nonumber\\&
    = (1+\lambda\Delta_{i})\nabla F_i(\mathbf{w}_t^i),
\end{align}

where $\Delta_{i}=F_i(\mathbf{w}_t^i)-F(\mathbf{w}_t^i)$.

The stochastic gradient descent by using Equation (\ref{eq:grad_Hi_batch}) as show:
\begin{align}
    \mathbf{w}_{t+1}^i = \mathbf{w}_t^i - \eta(1+\lambda\Delta_{i})\nabla F_i(\mathbf{w}_t^i)
    =\mathbf{w}_t^i - \eta_i\nabla F_i(\mathbf{w}_t^i),
\end{align}
where $\eta_i=\eta(1+\lambda\Delta_{i})$.

Therefore, our approach can be recognized as establishing an \textbf{dynamic learning rate} that allows the model update to achieve both performance and fairness objectives simultaneously. Based on these principles, we introduce the \premethod algorithm, as describe in Algorithm \ref{alg:method_without_DP}:

\begin{enumerate}
    \item \textbf{Server (lines \ref{alg0:line_server_for1}-\ref{alg0:line_server_lossAggre}):} The server executes $T$ rounds of communication, each round broadcasting the aggregated model $\mathbf{w}_{t}$ and the loss $F(\mathbf{w}_{t})$ to clients.
    \item \textbf{Local (lines \ref{alg0:line_local_for4}-\ref{alg0:line_local_update}):} For each batch $b$, we compute the dynamic learning rate $\eta_i$ in \textbf{lines \ref{alg0:line_local_delta}-\ref{alg0:line_local_etai}}. For \textbf{line \ref{alg0:line_local_update}}, we execute gradient descent by using $\eta_i$.
\end{enumerate}

\section{\method: Further Equipping \premethod with Differential Privacy}\label{sec:method}

\begin{figure*}[tbh]  
  \begin{center}
  \includegraphics[width=1.0\textwidth]{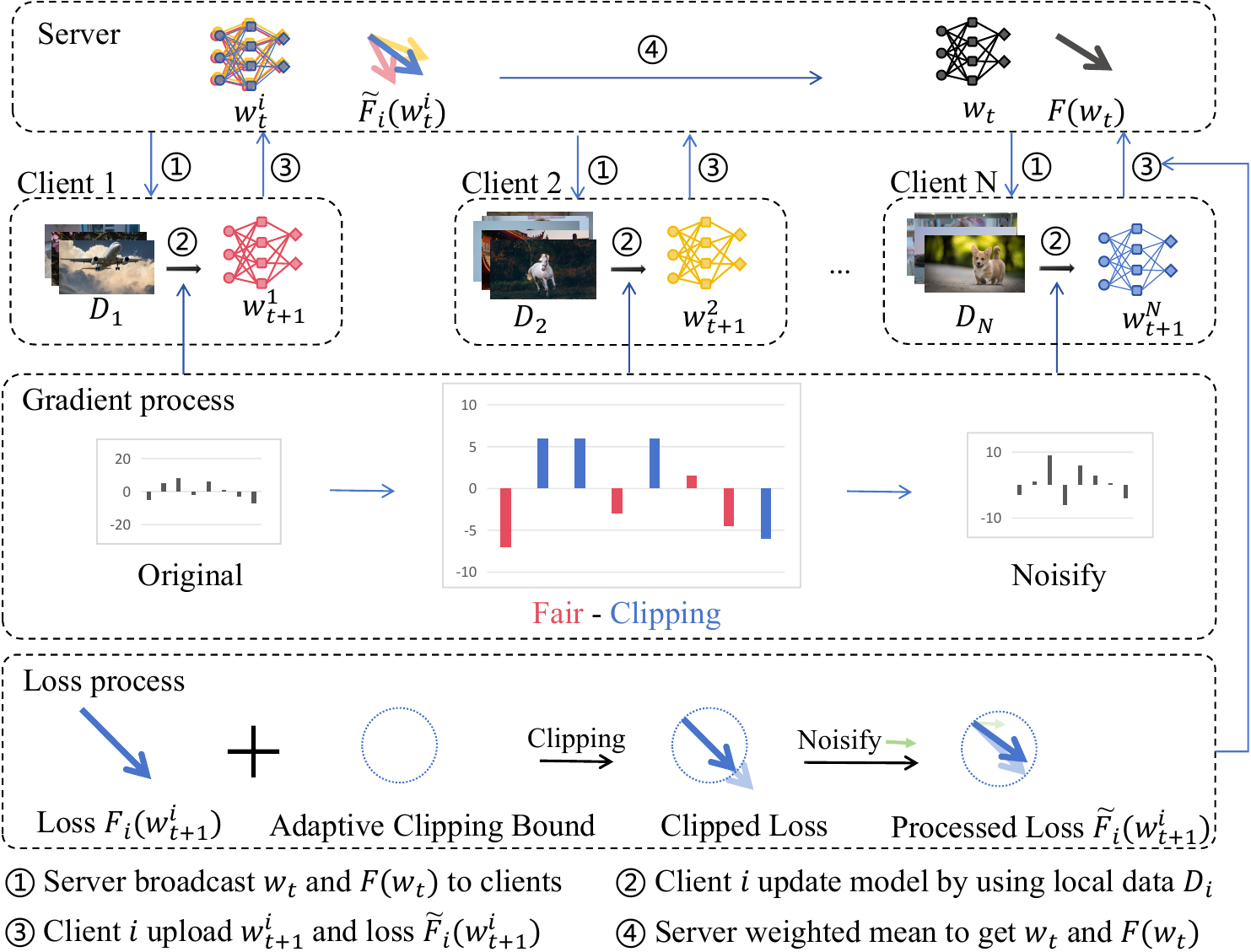}  
  \caption{\method framework. Besides the privately computed local model $\mathbf{w}_{t+1}^i$ from the ``Fair-Clipping'', client $i$ is also required to upload the private loss $\tilde{F}_i(\mathbf{w}_{t+1}^i)$. }  
  \label{fig:FL}  
  \end{center}
\end{figure*}

In this section, we present our algorithm, \method, which further integrates \premethod with differential privacy. The flow of \method algorithm is shown in Figure \ref{fig:FL}. In addition to adding Gaussian noise to the gradients, we introduce two additional processes to balance fairness and DP: 1) A fair-clipping strategy is added to \premethod. This strategy not only achieves sensitivity but also allows adjustments to the level of fairness. 2) An adaptive clipping method is employed when protecting the loss values that clients share with differential privacy, aiming to maximize utility.    
Next, we discuss the details of these two strategies.

\subsection{Fair-clipping Strategy for Gradient}\label{subsec:fair_clip}

In the general DPSGD algorithm \cite{Abadi2016DeepLearningWithDP,fu2023dpsur,BuZhiqi2022AutomaticClippingDPDLME,YangXiaodong2022NormalizedClippedSGD}, per-sample clipping is an essential process for obtaining sensitivity. In order to be compatible with per-sample clipping,  initially, we expand our objective function of client $i$ in \premethod, Equation (\ref{eq:fair_obj}), into Equation (\ref{eq:obj_H}):

\begin{equation}\label{eq:obj_H}
    \min_{\mathbf{w}_t^i} H_i(\mathbf{w}_t^i,\xi_j) = F_i(\mathbf{w}_t^i,\xi_j) + \frac{\lambda}{2} \left(F_i(\mathbf{w}_t^i,\xi_j)-F(\mathbf{w}_t^i) \right)^2 ,
\end{equation}

where $\xi_j \in \mathcal{B}_t^i$, $j=\{1,2,\cdots,|\mathcal{B}_i|\}$, and $|\mathcal{B}_i|$ is the batch size of client $i$.

Next, we compute the gradient of Equation (\ref{eq:obj_H}):

\begin{align}\label{eq:grad_of_H}
\nabla H_i(\mathbf{w}_t^i,\xi_j) 
= \left( 1+ \lambda \cdot \Delta_{i}^j  \right) \cdot \nabla F_i(\mathbf{w}_t^i,\xi_j) ,
\end{align}

where $\Delta_{i}^j= F_i(\mathbf{w}_t^i,\xi_j)-F(\mathbf{w}_t^i)  $.

To ensure the algorithm is safeguarded by differential privacy, we replace the gradient of the loss function with Equation (\ref{eq:grad_of_H}) and reformulate the gradient descent computation for client $i$ as follows:

\begin{align}
    \mathbf{w}_{t+1}^{i}
    &=\mathbf{w}_t^{i} - \eta \mathbf{\tilde{g}}^{i}_t \nonumber\\
    &= \mathbf{w}_t^{i} - \frac{\eta}{|\mathcal{B}_i|}\left[
    \sum^{|\mathcal{B}_i|}_{j=1} \frac{ \left(1+\lambda\cdot\Delta_{i}^j\right)\nabla F_i(\mathbf{w}_t^i,\xi_j) }{\max\left(1,\frac{\left(1+\lambda\cdot\Delta_{i}^j\right)\|\nabla F_i(\mathbf{w}_t^i,\xi_j)\|}{C}\right)}
    + \sigma C \cdot \mathcal{N}(0,\mathbf{I})
    \right]\nonumber\\&
    = \mathbf{w}_t^{i} - \frac{\eta}{|\mathcal{B}_i|}\left[
    \sum^{|\mathcal{B}_i|}_{j=1} C^{i,j}_t \cdot \nabla F_i(\mathbf{w}_t^i,\xi_j)
    + \sigma C \cdot \mathcal{N}(0,\mathbf{I})\right]\,,\label{eq:sgd_dynamic_clipping}
\end{align}
where:
\begin{equation}\label{eq:C_ijt}
    C^{i,j}_t = \min\left( 1+\lambda\cdot\Delta_{i}^j,\frac{C}{\| \nabla F_i(\mathbf{w}_t^i,\xi_j) \|} \right).
\end{equation}

Hence, our strategy can be considered as a \textbf{fair-clipping} technique, tailored to meet the demands of differential privacy while also nurturing fairness within the model. When $\lambda = 0$, Equation (\ref{eq:C_ijt}) corresponds to the traditional DP clipping method without fairness. For large $C$, Equation (\ref{eq:C_ijt}) simplifies to the first term, prompting \method to adjust towards enhanced fairness. Conversely, for small $C$, Equation (\ref{eq:C_ijt}) simplifies to the second term, resulting in the gradient being clipped to the bound $C$. The impact of $C$ on the performance of \method will be further discussed in the experiments. 

\subsection{Adaptive Clipping Method for Loss}\label{subsec:dp_loss}

Since calculating $\Delta_{i}^j$ requires clients to upload $F_i(\mathbf{w}_t^i)$, differential privacy must be incorporated to ensure the algorithm preserves privacy. As shown in lines~\ref{alg1:line_local_for4}-\ref{alg1:line_local_dp_loss} of Algorithm~\ref{alg:method}, $F_i(\mathbf{w}_t^i)$ must be clipped to the interval $[0, C_l^{i,t}]$, followed by the addition of Gaussian noise with a mean of 0 and a standard deviation of $\sigma_l \cdot C_l^{i,t}$ to ensure differential privacy. However, determining an appropriate clipping bound $C_l^{i,t}$ for $F_i(\mathbf{w}_t^i)$ is challenging. In federated learning, due to the heterogeneity of data held by different clients, $F_i(\mathbf{w}_t^i)$ can vary significantly. A clipping norm that is too large introduces excessive noise, while one that is too small leads to aggressive clipping of gradient directions, impairing model performance. Therefore, an adaptive clipping strategy is necessary.

Based on the data heterogeneity of federated learning, we propose an adaptive clipping method for $F_i(\mathbf{w}_t^i)$. For the $i$-th client, the adaptive clipping method uses the \textit{differentially private mean of the previous round} as the clipping bound for the current round. The clipping bound $C_l^{i,t}$ for round $t$ and client $i$ is defined as follows, when $F_i(\mathbf{w}_{t-1}^i)$ is the individual loss of the previous round and $\sigma^{t-1}_l$ is the individual noise multiplier of the previous round.

\begin{equation} \label{equ:adaclip1}
C_l^{i,t} = \frac{\sum_{j=1}^{|\mathcal{B}_i|} \operatorname{clip}\left(F_i(\mathbf{w}_{t-1}^i,\xi_j)\right)+\mathcal{N}\left(0, ({C_l^{i,t-1}} \cdot {\sigma_l})^2 \right)}{|\mathcal{B}_i|},
\end{equation}
where
\begin{equation} \label{equ:adaclip2}
\operatorname{clip}\left(F_i(\mathbf{w}_{t-1}^i,\xi_j)\right)= \min({C_l^{i,t-1}},\max(0,F_i(\mathbf{w}_{t-1}^i,\xi_j))).
\end{equation}

\begin{algorithm}[!t]
\caption{\method}\label{alg:method}
\KwIn{loss function $F(\mathbf{w})$, learning rate $\eta$, noise multiplier for gradient $\sigma$, noise multiplier for loss $\sigma_l$, fairness hyperparameter $\lambda$, batch sample ratio $q$, original clipping bound $C$.}
\KwOut{the final trained model $\mathbf{w}_T$}
\Retract{\textbf{SERVER}}{\label{alg1:line_server}
\For{$t = 0,1,\cdots,T-1$ }
{\label{alg1:line_server_for1}
\For{$i = 1,2,\cdots,N$ parallel }{\label{alg1:line_server_for2}
$\mathbf{w}_{t+1}^i,\tilde{F}_i(\mathbf{w}_{t+1}^i)=\text{LOCAL}\left( \mathbf{w}_{t},F(\mathbf{w}_{t})\right)$\label{alg1:line_server_local_update}\;
}
$\mathbf{w}_{t+1} = \sum_{i=1}^N p_i \mathbf{w}_{t+1}^i$\label{alg1:line_server_modelAggre}\;
$F(\mathbf{w}_{t+1}) = \sum_{i=1}^N p_i \tilde{F}_i(\mathbf{w}_{t+1}^i)$\label{alg1:line_server_lossAggre}\;
}
\Return $\mathbf{w}_T$\label{alg1:line_server_return}\;
}
\Retract{\textbf{LOCAL}}{\label{alg1:line_local}
$\text{Download } \mathbf{w}_t, F(\mathbf{w}_t) \text{ and } \mathbf{w}_t^{i} \leftarrow \mathbf{w}_t$\label{alg1:line_local_download}\;
Sample randomly a batch $\mathcal{B}_i$ with probability $q$\label{alg1:line_local_sample}\;
\For{$j = 1,2,\cdots,|\mathcal{B}_i|$ }
{\label{alg1:line_local_for3}
$\Delta_{i}^j= F_i(\mathbf{w}_t^i,\xi_j)-F(\mathbf{w}_t)$\label{alg1:line_local_delta}\;
Compute $C^{i,j}_t$ by Equation (\ref{eq:C_ijt})\label{alg1:line_local_Cijt}\;
$\mathbf{\tilde{g}}_t^{i,j}=C_t^{i,j}\cdot\nabla F_i(\mathbf{w}_t^i,\xi_j)$\label{alg1:line_local_grad}\;
}
$\mathbf{w}_{t+1}^i=\mathbf{w}_t^{i} - \frac{\eta}{|\mathcal{B}_i|} \left( \sum_{j=1}^{|\mathcal{B}_i|} \mathbf{\tilde{g}}_t^{i,j} + \sigma C \cdot \mathcal{N}(0,\mathbf{I})\right)$\label{alg1:line_local_update}\;

Compute clipping bound of loss $C_l^{i,t}$ by Equation~(\ref{equ:adaclip1}) and (\ref{equ:adaclip2}) \; \label{alg1: adadclip}
\For{$j =1,2,\cdots,|\mathcal{B}_i|$ }
{\label{alg1:line_local_for4}
Compute $F_i(\mathbf{w}_{t+1}^i,\xi_j)$\label{alg1:line_local_loss}\;

$f_{i,j}=\min({C_l^{i,t}},\max(0,F_i(\mathbf{w}_{t+1}^i,\xi_j)))$\; \label{alg1:line_19} 
}
$\tilde{F}_i(\mathbf{w}_{t+1}^i) = \frac{1}{|\mathcal{B}_i|} \left(\sum_{j=1}^{|\mathcal{B}_i|} f_{i,j} + \sigma_l C_l^{i,t} \mathcal{N}(0,\mathbf{I})\right)$\label{alg1:line_local_dp_loss}\; \label{alg1:line_20}
\Return $ \mathbf{w}_{t+1}^i, \tilde{F}_i(\mathbf{w}_{t+1}^i)$\label{alg1:line_local_return}\;
}
\end{algorithm}

To ensure differential privacy during the clipping process, we use the noise-added loss $C_l^{i,t}$ as the clipping bound. For each communication round, the clipping bound follows the post-processing property of differential privacy.

\subsection{Overall Algorithm} \label{sub:overall}

We utilize the equation form Equation (\ref{eq:sgd_dynamic_clipping}) to enhance Algorithm \ref{alg:method}. Compared with the Algorithm \ref{alg:method_without_DP}, \method solely necessitates the application of differential privacy to gradient and loss.

\begin{enumerate}
    

    \item \textbf{DPSGD (lines \ref{alg1:line_local_for3}-\ref{alg1:line_local_update}):} In \textbf{lines \ref{alg1:line_local_delta}-\ref{alg1:line_local_grad}}, we compute the fair-clipping bound $C_t^{i,j}$ and the per-sample gradient $\nabla F_i(\mathbf{w}_{t}^i,\xi_j)$ to get the processed gradient $\mathbf{\tilde{g}}_t^{i,j}$. Then, we add mean 0 noise to the sum of per-sample processed gradient $\mathbf{\tilde{g}}_t^{i,j}$ in \textbf{line \ref{alg1:line_local_update}} and execute the gradient decent.
    \item \textbf{DP for local loss (lines \ref{alg1: adadclip}-\ref{alg1:line_20}):} As described in Section \ref{subsec:dp_loss}, add noise to the clipped local loss by using adaptive clipping bound $C_l^{i,t}$.
\end{enumerate}

\section{Finding Optimal Fairness Parameter}
\label{sec:guarantees}

In this section, we discuss how to identify an optimal fairness parameter $\lambda$ through the convergence analysis of the \method algorithm. 

All assumptions and detailed procedures for the convergence analysis are provided in Appendix~\ref{append:proof_of_ca}, leading to the following result:

\begin{theorem}[Convergence rate of \method\, (Simplified version of Theorem~\ref{theo:ca})]
\begin{align}\label{eq:simp_ca_result}
\mathbb{E}[F(\mathbf{w}_T)] - F^* &\leq 
\mathcal{O}\left( \frac{L G^2 \lambda^2}{\mu^2 T} \right) +
\mathcal{O}\left( \frac{L^2 \Gamma \lambda}{\mu^2 T} \right) 
\nonumber\\&\quad +
\mathcal{O}\left( \frac{L \sigma^2 C^2 d}{\mu^2 T \lambda} \right) + 
\mathcal{O}\left( \frac{ L \mathbb{E}\| \mathbf{w}_1 - \mathbf{w}^* \|^2}{ T} \right)\,,
\end{align}
where $L$, $\mu$, and $G$ are parameters defined in Assumptions~\ref{ass:l_smooth}--\ref{ass:grad_bound}, and $\Gamma$ is a measure of heterogeneity; see Appendix~\ref{subsec:ca} for details.
\end{theorem}

As demonstrated in \eqref{eq:simp_ca_result}, the fairness parameter $\lambda$ ($\lambda>0$) exhibits convex influence on the convergence rate. The analytical solution of this relationship will be obtained via extremum analysis in Section~\ref{subsec:optimal_lambda}.

\subsection{Optimal Fairness Parameter $\lambda^*$}\label{subsec:optimal_lambda}







Assuming $\Delta_{i}^j$ is bounded as $Q_0\leq \Delta_{i}^j \leq Q_1$, then $1+Q_0 \lambda \leq C_t^{i,j}\leq 1+Q_1 \lambda$. We rearrange the convergence upper bound in Equation (\ref{eq:simp_ca_result}) to obtain a function in terms of $\lambda$:

\begin{equation}
    P(\lambda)=\frac{L}{2\mu T}\frac{a_1\lambda^3+a_2\lambda^2+a_3\lambda+a_4}{a_5\lambda+1},
\end{equation}

where $a_1=G^2Q_1^3$, $a_2=6G^2Q_1^2$, $a_3=9Q_1G^2+2L\Gamma Q_1+2Q_0\mathbb{E}\|\mathbf{w}_1-\mathbf{w}^*\|^2$, $a_4=4G^2+2L\Gamma+\frac{2\sigma^2C^2d}{\hat{B}^2}+\mathbb{E}\|\mathbf{w}_1-\mathbf{w}^*\|^2$, $a_5 = 2Q_0$.


Let $\mathcal{F}(\lambda)=\frac{a_1 \lambda^3+a_2 \lambda^2+a_3\lambda +a_4}{a_5\lambda+1}$, then $P(\lambda)=\frac{L}{2\mu T}\mathcal{F}(\lambda)$, where $\frac{L}{2\mu T}$ can be considered a constant term. Obviously, when $\mathcal{F}(\lambda)$ is minimized, $P(\lambda)$ is also minimized. The derivative of $\mathcal{F}(\lambda)$ is as follows:

\begin{align}
    \mathcal{F}'(\lambda) 
    = \frac{2a_1a_5\lambda^3+(a_2a_5+3a_1)\lambda^2+2a_2\lambda+a_3-a_4a_5}{(a_5\lambda+1)^2}.
\end{align}

Let the denominator of \(\mathcal{F}'(\lambda)\) be a new function:
\begin{equation}
    \mathcal{G}(\lambda)=2a_1a_5\lambda^3+(3a_1+a_2a_5)\lambda^2+2a_2\lambda+a_3-a_4a_5.
\end{equation}
Since $(a_5\lambda+1)^2 > 0$, the sign of $\mathcal{F}'(\lambda)$ is consistent with that of $\mathcal{G}(\lambda)$.

Using the discriminant of a cubic function $\Delta = b^2-3ac$ (a general cubic function is expressed as $a\lambda^3+b\lambda^2+c\lambda+d$),
\begin{align}
    \Delta &= (3a_1+a_2a_5)^2 - 3(2a_1a_5\cdot 2a_2)
    \nonumber\\&
    = 9a_1^2+a_2^2a_5^2+6a_1a_2a_5 - 12a_1a_2a_5\nonumber\\
    &=(3a_1-a_2a_5)^2 \ge 0.
\end{align}

The derivative of $\mathcal{G}(\lambda)$ is given by
\begin{equation}
    \mathcal{G}'(\lambda) = 6a_1a_5\lambda^2+2(a_2a_5+3a_1)\lambda+2a_2.
\end{equation}
It is evident that when $\lambda > 0$, $\mathcal{G}'(\lambda) > 0$.

If $\mathcal{F}(\lambda)$ has an extremum, then $\mathcal{F}'(\lambda)$ has a root, meaning $\mathcal{G}(\lambda)$ has a root. Substituting $\lambda=0$, we have $\mathcal{G}(0) = a_3-a_4a_5$, evidently when $a_3<a_4a_5$, $\mathcal{G}(\lambda)$ has a root.

Let $\lambda = x - \frac{a_2a_5+3a_1}{6a_1a_5}$, replacing $\lambda$ with $x$ in $\mathcal{G}(\lambda)$ yields

\begin{equation}
    \mathcal{G}(x) = x^3+a_6x+a_7,
\end{equation}
where 
\begin{itemize}
    \item $a_6=\frac{-(3a_1-a_2a_5)^2}{12a_1^2a_5^2}$,
    \item $a_7=\frac{108a_1^2a_5^2(a_3-a_4a_5)-36(a_1a_2a_5)(3a_1+a_2a_5)+2(3a_1+a_2a_5)^3}{216a_1^3a_5^3}$.
\end{itemize}

Using the root-finding formula to calculate $\mathcal{G}(x)=0$ gives the solutions $x_1,x_2,x_3$.

\begin{equation*}
    x_1 = \sqrt[3]{-\frac{a_7}{2} + \sqrt{\frac{a_7^2}{4} + \frac{a_6^3}{27}}} + \sqrt[3]{-\frac{a_7}{2} - \sqrt{\frac{a_7^2}{4} + \frac{a_6^3}{27}}},
\end{equation*}

\begin{equation*}
x_2 = \omega \sqrt[3]{-\frac{a_7}{2} + \sqrt{\frac{a_7^2}{4} + \frac{a_6^3}{27}}} + \omega^2 \sqrt[3]{-\frac{a_7}{2} - \sqrt{\frac{a_7^2}{4} + \frac{a_6^3}{27}}},
\end{equation*}

\begin{equation*}
x_3 = \omega^2 \sqrt[3]{-\frac{a_7}{2} + \sqrt{\frac{a_7^2}{4} + \frac{a_6^3}{27}}} + \omega \sqrt[3]{-\frac{a_7}{2} - \sqrt{\frac{a_7^2}{4} + \frac{a_6^3}{27}}}.
\end{equation*}

where imaginary number $\omega = \frac{-1+\sqrt{3}\mathbf{i}}{2}$, $\lambda_{l}=x_{l}-\frac{a_2a_5+3a_1}{6a_1a_5}$, $l\in \{1,2,3\}$. 

Furthermore, since $\lambda_1\lambda_2\lambda_3 = -\frac{a_3-a_4a_5}{2a_1a_5} > 0$, $\lambda$ has only one positive real root. If the equation has $\mathrm{m}$ real roots, the optimal solution for $\lambda$ is $\lambda^* = \max_{\mathrm{m}}(\lambda_\mathrm{m})$.

In conclusion, $\mathcal{F}'(\lambda) \ge 0 \iff G(\lambda) \ge 0 \iff \lambda \ge \lambda^*$, therefore $\mathcal{F}(\lambda)$ has a minimum value $\mathcal{F}(\lambda^*)$, and $P(\lambda)$ has a minimum value $P(\lambda^*)$.


\section{Privacy Analysis} \label{sec:privacy_analysis}

At the beginning of privacy analysis, we clarify that: (1) FedFDP does not assume Secure Aggregation (SecAgg)\cite{bonawitz2017practical}, allowing the server to observe individual client uploads; (2) we guarantee \textit{sample-level DP} via per-sample clipping to protect individual data records; and (3) SecAgg is orthogonal to our method—while it hides per-client updates, it cannot replace DP in preventing the aggregated model from memorizing sensitive data.

As Algorithm~\ref{alg:method} shown, the $i$-th client will return model parameters $\mathbf{w}_{t+1}^i$ and loss value $\tilde{F}_i(\mathbf{w}_{t+1}^i)$ before the server aggregates the model parameters in round $t+1$. $\mathbf{w}_{t+1}^i$ and $\tilde{F}_i(\mathbf{w}_{t+1}^i)$ both access the private training data of client $i$, so they are both to perform differential privacy preservation, as in lines~\ref{alg1:line_local_for3}-\ref{alg1:line_local_grad} and lines~\ref{alg1:line_local_for4}-\ref{alg1:line_local_dp_loss} of the Algorithm~\ref{alg:method}, respectively. We will analyze their privacy with RDP separately and finally combine their privacy loss to get the overall privacy loss of \method algorithm.

\subsection{Privacy Loss of $\mathbf{w}_{t+1}^i$}
\begin{theorem}\label{the:rdp-of-dpsgd}
After $T$ rounds local updates, the RDP of the $\mathbf{w}_{T}^i$ in $i$-th client satisfies:

\begin{equation}
    R_{model}^i(\alpha) = \frac{T}{\alpha-1}\sum_{k=0}^{\alpha}\left(\begin{array}{l}
\alpha \\ k
\end{array}\right)(1-q)^{\alpha-k} q^{k} \exp \left(\frac{k^{2}-k}{2 \sigma^{2}}\right),
\end{equation}
where  $\sigma$ is noise multiplier of the $\mathbf{w}_{t+1}^i$, and $\alpha > 1$ is the order.
\end{theorem}
\textbf{Proof.} We will prove Theorem \ref{the:rdp-of-dpsgd} in the following two steps: (i) use the
RDP of the sampling Gaussian mechanism to calculate the privacy
cost of each model update, and (ii) use the composition of RDP mechanisms to compute the privacy cost of multiple model updates.

\begin{definition}\label{privacy RDP privacy budget of SGM}
(RDP privacy budget of SGM\cite{IlyaMironov2019RnyiDP}). Let $SG_{q,\sigma}$, be the Sampled Gaussian Mechanism for some function $f$. If $f$ has sensitivity $1$, $SG_{q,\sigma}$ satisfies $(\alpha,R)$-RDP whenever
\begin{equation}
R \leq \frac{1}{\alpha-1} \log \max(A_{\alpha}(q,\sigma),B_{\alpha}(q,\sigma))\,,
\end{equation}
where
\begin{equation}
\left\{\begin{array}{l}
A_{\alpha}(q, \sigma) = \mathbb{E}_{z \sim \vartheta_{0}}\left[\left(\vartheta(z) / \vartheta_{0}(z)\right)^{\alpha}\right] \,,\\
B_{\alpha}(q, \sigma) = \mathbb{E}_{z \sim \vartheta}\left[\left(\vartheta_{0}(z) / \vartheta(z)\right)^{\alpha}\right]\,.
\end{array}\right.
\end{equation}
with $\vartheta_{0} = \mathcal{N}\left(0, \sigma^{2}\right), \vartheta_{1} = \mathcal{N}\left(1, \sigma^{2}\right) \mbox { and } \vartheta =(1-q) \vartheta_{0}+q \vartheta_{1}$\,.

\end{definition}

Further, it holds for $\forall(q,\sigma)\in(0,1], \mathbb{R}^{+*},A_{\alpha}(q,\sigma) \geq B_{\alpha}(q, \sigma) $. Thus, $ S G_{q, \sigma} $ satisfies $ \left(\alpha, \frac{1}{\alpha-1} \log \left(A_{\alpha}(q, \sigma)\right)\right)$-RDP .

Finally, the existing work \cite{IlyaMironov2019RnyiDP} describes a procedure to compute $A_{\alpha}(q,\sigma)$ depending on integer $\alpha$:
\begin{equation}
A_{\alpha}=\sum_{k=0}^{\alpha}\left(\begin{array}{l}
\alpha \\ k
\end{array}\right)(1-q)^{\alpha-k} q^{k} \exp \left(\frac{k^{2}-k}{2 \sigma^{2}}\right)\,.
\end{equation}

\begin{definition} \label{privacy Composition of RDP}
(Composition of RDP\cite{IlyaMironov2017RnyiDP}). For two randomized mechanisms $f, g$ such that $f$ is $(\alpha,R_1)$-RDP and $g$ is $(\alpha,R_2)$-RDP the composition of $f$ and $g$ which is defined as $(X, Y )$(a sequence of results), where $ X \sim f $ and $Y \sim g$, satisfies $(\alpha,R_1+R_2)-RDP$
\end{definition}

From Definition~\ref{privacy RDP privacy budget of SGM} and  Definition~\ref{privacy Composition of RDP}, the Theorem~\ref{the:rdp-of-dpsgd} is obtained.

\subsection{Privacy Loss of $\tilde{F}_i(\mathbf{w}_{t+1}^i)$}
\begin{theorem}\label{the:rdp-of-validation}
After $T$ rounds local updates, the RDP of the $\tilde{F}_i(\mathbf{w}_{t+1}^i)$ in $i$-th client satisfies:

\begin{equation}
    R_{loss}^i(\alpha) = \frac{T}{\alpha-1}\sum_{k=0}^{\alpha}\left(\begin{array}{l}
\alpha \\ k
\end{array}\right)(1-q)^{\alpha-k} q^{k} \exp \left(\frac{k^{2}-k}{2 \sigma_{l}^{2}}\right),
\end{equation}    
where  $\sigma_l$ is noise multiplier of the $\tilde{F}_i(\mathbf{w}_{t+1}^i)$, and $\alpha > 1$ is the order.
\end{theorem}

The proof is similar to that of the $\mathbf{w}_{t+1}^i$, so we omit it.

\subsection{Privacy Loss of \method}
Since both $\mathbf{w}_{T}^i$ and $\tilde{F}_i(\mathbf{w}_{t+1}^i)$ access the training set, we need to combine their RDP sequentially using Definition~\ref{privacy Composition of RDP}, and then use Lemma~\ref{lem:conversion2} to convert it to $(\epsilon,\delta)$-DP. Lastly, we can get the Privacy loss of \method as follows.

\begin{theorem}\label{the:privacy-loss-FedPDF}
	({\bf Privacy loss of \method}). The privacy loss in $i$-th client of \method satisfies:
		\begin{align}
		\begin{split}
			\begin{aligned}
				(\epsilon^i,\delta^i)=&(R_{model}^i(\alpha)+ R_{loss}^i(\alpha) +\ln ((\alpha-1) / \alpha) \\
				&-(\ln \delta+ \ln \alpha) /(\alpha-1),\delta),
			\end{aligned}
		\end{split}
	\end{align}
where $0<\delta<1$, $R_{model}^i(\alpha)$ is the RDP of $w_{t+1}^i$ is computed by Theorem~\ref{the:rdp-of-dpsgd}, and $R_{loss}^i(\alpha)$ is the RDP of $\tilde{F}_i(\mathbf{w}_{t+1}^i)$ which is computed by Theorem~\ref{the:rdp-of-validation}.
\end{theorem}

\begin{lemma}\label{lem:conversion2}
({\bf Conversion from RDP to DP~\cite{balle2020hypothesis}}). if a randomized mechanism $f : \mathcal{X}^n \rightarrow \mathbb{R}^d$  satisfies $(\alpha,R)$-RDP , then it satisfies$(R+\ln ((\alpha-1) / \alpha)-(\ln \delta+ \ln \alpha) /(\alpha-1), \delta)$-DP for any $0<\delta<1$.
\end{lemma}



\section{Experiments} \label{sec:experiments}
In this section, we demonstrate the effectiveness of \premethod/\method through answering the following three research questions:
\begin{itemize}
    \item \textbf{RQ1} How effective is \premethod~/ \method under classification tasks with real-world datasets?
    \item \textbf{RQ2} How robust is \premethod~/ \method across different experimental settings?
    \item \textbf{RQ3} What is the performance of \method under differential hyper-parameters?
\end{itemize}


To address the aforementioned three RQs, Section~\ref{subsec:default_settings} introduces the default experimental settings used in this study, including the description of baselines and the configuration of the experimental code. Section~\ref{subsec:exp_1} presents comparative experiments conducted on three datasets under default parameters involving \premethod, \method, and six baseline methods to answer RQ1. Section~\ref{subsec:exp_2} addresses RQ2 through experiments on heterogeneity and scalability. Finally, Section~\ref{subsec:exp_3} performs ablation studies on $\lambda$, $C$, $\sigma$, $\epsilon$, and the target accuracy to answer RQ3.

\subsection{Default Settings}\label{subsec:default_settings}
Prior to addressing the three research questions, an outline of the default experimental configuration is provided. 

\textbf{Baselines.} We compared our algorithms against six baseline methods. For \premethod, the baseline algorithms were left unmodified; for \method, DP was applied to each baseline algorithm to enable a fair comparison, which include FedAvg\cite{mcmahan2017communication}, SCAFFOLD\cite{karimireddy2020scaffold}, FedProx\cite{li2020fedprox}, FedDyn\cite{acar2021feddyn}, ALI-DPFL\cite{ling2024ali} and q-FFL\cite{TianLi2020qFFL}.

\begin{table*}[t]
\centering
\caption{Comparison of Communication Overhead and Transmission Requirements. $d$ represent the dimension of model.}
\label{tab:comm_overhead}
\begin{tabular}{l|c|c|l}
\hline
\textbf{Algorithm} & \textbf{\begin{tabular}[c]{@{}c@{}}Comm. Cost\\ (Per Round)\end{tabular}} & \textbf{\begin{tabular}[c]{@{}c@{}}Extra Transmitted\\ Variables\end{tabular}} & \textbf{Description of Overhead} \\ \hline
FedAvg & $\mathcal{O}(d)$ & None & Transmits only model parameters/gradients. \\ \hline
FedProx & $\mathcal{O}(d)$ & None & Adds proximal term locally. \\ \hline
q-FFL & $\mathcal{O}(d)$ & Scalar & Sends scalar loss for re-weighting. \\ \hline
SCAFFOLD & $\mathcal{O}(2d)$ & Control Variates & Transmits both model and control variates. \\ \hline
FedDyn & $\mathcal{O}(d)$ & None & Improve SCAFFOLD without control variables. \\ \hline
ALI-DPFL & $\mathcal{O}(d)$ & Scalar & Server broadcast scalar local steps to clients. \\ \hline
FedFDP & $\mathcal{O}(d)$ & Scalar & Transmits scalar loss for fairness/privacy. \\ \hline
\end{tabular}
\end{table*}
\textbf{Communication Overhead Analysis.} Table \ref{tab:comm_overhead} presents a detailed comparison of the communication complexity per round for FedFDP and other state-of-the-art baselines. While methods like SCAFFOLD introduce control variates to mitigate client drift, they double the communication payload to $\mathcal{O}(2d)$, which can be prohibitive in bandwidth-constrained wireless networks. In contrast, FedFDP maintains a communication complexity of $\mathcal{O}(d)$, aligning with efficient baselines such as FedAvg and FedDyn . Although FedFDP incorporates a fairness-aware mechanism and differential privacy, the additional transmission overhead is restricted to a negligible scalar value (representing the fairness loss), similar to the approach in q-FFL and ALI-DPFL. This demonstrates that FedFDP achieves robust privacy and fairness without compromising communication efficiency.

\textbf{Tasks setting.}  By presenting the comparative experimental results of \premethod and \method with multiple baseline methods on three datasets: MNIST \cite{Yann1998MNIST}, FashionMNIST \cite{xiao2017fashionMNIST}, and CIFAR10 \cite{AlexKrizhevsky2009LearningML}, we adopted a widely used heterogeneous settings~\cite{TaoLin2020Ensemble,li2021model} to 10 clients which controlled by a Dirichlet distribution denoted as $\mathrm{Dir}(\beta)$, and the default value of $\beta=0.1$~\cite{TaoLin2020Ensemble,WangJ2020FedNova}. We use a 4-layer CNN architecture \cite{mcmahan2017communication} which consists of two convolutional layers and two fully connected layers as model architecture. 

\textbf{Implementation environment.} We implement our experiment using PyTorch-1.8 and run all experiments on a server with one Intel i9 13900ks CPU (24 cores), 64GB of memory, and one NVIDIA 4090 GPU, running on Windows 10. 

\textbf{Hyperparameters}. For \premethod and its baseline algorithms, we set learning rate $\eta=0.1$. For \method and its baseline algorithms, we set the $\eta=1.0$, the batch sample ratio $q=0.05$, clipping bound for gradient $C=0.1$, the noise multipliers for gradient $\sigma=2.0$, the privacy budget $\epsilon=3.52$ and the $\delta=1.0 \times 10^{-5}$. In particular, for the loss values in the \method algorithm that additionally require differential privacy processing, we set the clipping bound for loss $C_{l}=2.5$ and the noise multipliers for loss $\sigma_{l}=5.0$.


\subsection{Effective in Real-world Datasets}\label{subsec:exp_1}

To address \textbf{RQ1}, we conducted experiments on three datasets for both \premethod and \method under the default parameter settings described in Section~\ref{subsec:default_settings}. The results, presented in Table~\ref{tab:exp1-1} and Table~\ref{tab:exp1-2}, report the test accuracy ($\%$) and fairness measure ($\Psi$).

\begin{table*}[!t]
  \centering
      \caption{\premethod and baselines: results of test accuracy (\%) and fairness ($\Psi$).}
    \label{tab:exp1-1}
  \resizebox{1.0\linewidth}{!}{
    \begin{tabular}{l|*{2}{c}|*{2}{c}|*{2}{c}}
    \toprule
    Datasets & \multicolumn{2}{c|}{MNIST} & \multicolumn{2}{c|}{FashionMNIST} & \multicolumn{2}{c}{CIFAR10} \\
    \midrule
   & Acc.($\uparrow$) & $\Psi(\downarrow)$ & Acc.($\uparrow$) & $\Psi(\downarrow)$ & Acc.($\uparrow$) & $\Psi(\downarrow)$   \\
    \midrule
    FedAvg & 98.67$\pm$0.09 & 7.3e-3$\pm$5.4e-3 & 87.05$\pm$1.02 & 3.6e-1$\pm$2.2e-1 & 62.03$\pm$1.00 & 1.1e0$\pm$8.9e-1 \\
    SCAFFOLD & 98.45$\pm$0.03 & 3.0e-2$\pm$2.6e-2 & 86.77$\pm$0.25 & 6.1e0$\pm$4.0e0 & 62.21$\pm$1.37 & 9.9e0$\pm$1.1e1 \\
    FedProx & 98.70$\pm$0.13 & 7.5e-3$\pm$5.1e-3 & 87.30$\pm$0.51 & 3.5e-1$\pm$2.1e-1 & 62.13$\pm$0.20 & 1.2e0$\pm$3.1e-1 \\
    FedDyn & 98.40$\pm$0.23 & 8.0e-3$\pm$2.1e-3 & 87.48$\pm$0.31 & 6.2e-1$\pm$3.5e-1 & 62.33$\pm$1.15 & 3.2e0$\pm$8.8e-1 \\
    q-FFL & 98.72$\pm$0.03 & 4.7e-3$\pm$1.0e-3 & 87.35$\pm$0.13 & 5.1e-1$\pm$5.2e-2 & \textbf{63.33$\pm$0.21} & 9.4e-1$\pm$3.3e-2 \\
    \premethod & \textbf{98.75$\pm$0.09} & \textbf{3.9e-3$\pm$1.3e-3} & \textbf{87.70$\pm$0.76} & \textbf{2.6e-1$\pm$2.3e-1} & 62.38$\pm$0.88 & \textbf{8.5e-1$\pm$6.6e-1} \\
    \bottomrule
    \end{tabular}
    }
\end{table*}

\begin{table*}[!t]
  \centering
      \caption{\method and baselines: results of test accuracy (\%) and fairness ($\Psi$), while the privacy budget $\epsilon=3.52$.}
    \label{tab:exp1-2}
  \resizebox{1.0\linewidth}{!}{
    \begin{tabular}{l|*{2}{c}|*{2}{c}|*{2}{c}}
    \toprule
        Datasets & \multicolumn{2}{c|}{MNIST} & \multicolumn{2}{c|}{FashionMNIST} & \multicolumn{2}{c}{CIFAR10} \\
    \midrule
    & Acc.($\uparrow$) & $\Psi(\downarrow)$ & Acc.($\uparrow$) & $\Psi(\downarrow)$& Acc.($\uparrow$) & $\Psi(\downarrow)$  \\
    \midrule
    FedAvg & 93.40$\pm$0.47 & 1.1e11$\pm$0.5e10 & 84.15$\pm$1.97 & 3.0e8$\pm$1.4e8 & 52.61$\pm$0.17 & 6.1e9$\pm$5.2e9 \\
    SCAFFOLD & 93.95$\pm$1.39 & 7.0e10$\pm$4.9e9 & 83.41$\pm$4.87 & 3.3e9$\pm$4.7e9 & 53.85$\pm$1.16 & 6.2e9$\pm$1.3e9 \\
    FedProx & 90.50$\pm$3.95 & 5.3e10$\pm$7.3e9 & 83.85$\pm$0.98 & 7.9e8$\pm$4.5e8 & 51.34$\pm$0.91 & 4.4e9$\pm$2.1e9 \\
    FedDyn & 91.42$\pm$2.25 & 5.6e10$\pm$9.8e9 & 84.04$\pm$1.21 & 8.8e8$\pm$3.2e8 & 52.14$\pm$1.39 & 5.5e9$\pm$1.2e9 \\
    ALI-DPFL & 90.89$\pm$1.65 & 3.3e10$\pm$8.3e9 & 83.65$\pm$1.68 & 6.6e8$\pm$3.5e8 & 52.05$\pm$1.26 & 3.6e9$\pm$1.6e9 \\
    q-FFL & 93.74$\pm$1.41 &7.8e10$\pm$1.1e11 & 83.13$\pm$2.74 & 4.2e9$\pm$2.6e9 & 48.46$\pm$1.00 & 4.7e9$\pm$1.9e9 \\
    \method & \textbf{95.13$\pm$0.83} & \textbf{2.3e10$\pm$1.0e10} & \textbf{85.99$\pm$0.76} & \textbf{2.8e8$\pm$0.8e8} & \textbf{54.21$\pm$0.98} & \textbf{2.6e9$\pm$1.6e9} \\
    \bottomrule
    \end{tabular}
    }
\end{table*}

Table \ref{tab:exp1-1} presents the experimental results of the \premethod algorithm compared to baseline methods. \premethod significantly enhances fairness while maintaining substantial model performance across the MNIST, FashionMNIST, and CIFAR10 datasets, with respective increases of 17.0\%, 25.7\%, and 9.6\%. A plausible explanation is that, as the training process progresses, \premethod increasingly prioritizes fairness, resulting in a reduction in $\eta_i$. Compared to training methods with a fixed learning rate, learning rate decay accelerate the convergence of gradient-based methods \cite{Martin2023OnlineConvexProgramming}.

Table~\ref{tab:exp1-2} presents the experimental results of the \method algorithm compared to baseline methods. In comparison with Table~\ref{tab:exp1-1}, the introduction of DP leads to a slight decrease in accuracy, approximately in the range of $3\%$\,-\,$10\%$. However, the fairness issue becomes more pronounced, with the corresponding metric increasing significantly in magnitude, indicating that DP exerts a substantial influence on fairness in FL. Under a fixed privacy budget of $\epsilon = 3.52$, \method significantly enhances fairness while matching the accuracy of the best-performing baseline, achieving improvements of $30.3\%$, $6.7\%$, and $27.8\%$ across three distinct datasets, respectively.

\subsection{Robust in Different Settings}\label{subsec:exp_2}

To answer \textbf{RQ2}, as shown in Table~\ref{tab:exp2-1} and Table~\ref{tab:exp2-2}, we extended the heterogeneity setting from the default $\mathrm{Dir}(0.1)$ to $\mathrm{Dir}(0.5)$ and $\mathrm{Dir}(1)$, and scaled the number of clients from the default 10 to 20 and 50 to examine scalability.


\begin{table*}[!t]
  \centering
      \caption{\premethod and baselines: results of test accuracy (\%) and fairness ($\Psi$) in different heterogeneity and scalability.}
    \label{tab:exp2-1}
  \resizebox{1.0\linewidth}{!}{
    \begin{tabular}{l|*{6}{c}|*{6}{c}}
    \toprule
    & \multicolumn{6}{c|}{Heterogeneity} & \multicolumn{6}{c}{Scalability}\\
    \midrule
    Datasets & \multicolumn{4}{c}{MNIST} & \multicolumn{2}{c|}{FashionMNIST} & \multicolumn{6}{c}{CIFAR10} \\
    \midrule
    Method & \multicolumn{2}{c}{$\mathrm{Dir}(0.1)$} & \multicolumn{2}{c}{$\mathrm{Dir}(0.5)$} & \multicolumn{2}{c|}{$\mathrm{Dir}(1)$}
    & \multicolumn{2}{c}{10 clients} & \multicolumn{2}{c}{20 clients} & \multicolumn{2}{c}{50 clients}\\
    \midrule
     & Acc.($\uparrow$) & $\Psi(\downarrow)$
     & Acc.($\uparrow$) & $\Psi(\downarrow)$
     & Acc.($\uparrow$) & $\Psi(\downarrow)$
     & Acc.($\uparrow$) & $\Psi(\downarrow)$
     & Acc.($\uparrow$) & $\Psi(\downarrow)$
     & Acc.($\uparrow$) & $\Psi(\downarrow)$
     \\
    \midrule
    FedAvg & 98.67 & 7.3e-3 & 98.37 & 1.2e-3 & 89.41 & 3.6e-2 & 62.03 & 1.1e0 & 60.89 & 1.5e0 & 59.78 & 1.5e0 \\
    SCAFFOLD & 98.45 & 5.0e-3 & 98.51 & 2.1e-3 & 86.39 & 3.4e-2 & 62.21 & 9.9e0 & 59.48 & 8.7e0 & 58.64 & 2.3e0 \\
    FedProx & 98.70 & 7.5e-3 & 98.35 & 1.2e-3 & 88.33 & 3.5e-2 & 62.13 & 1.2e0 & 60.33 & 2.2e0 & 61.31 & 1.6e0 \\
    FedDyn & 98.40 & 8.0e-3 & 98.36 & 1.8e-3 & 88.26 & 3.5e-2 & 62.33 & 9.4e-1 & 61.35 & 2.6e0 & 62.52 & 1.5e0 \\
    q-FFL & 98.72 & 4.7e-3 & 98.61 & \textbf{1.1e-3} & 89.03 & 3.6e-2 & \textbf{63.33} & 9.4e-1 & 60.63 & 1.3e0 & 59.65 & 1.5e0 \\
    \premethod & \textbf{98.75} & \textbf{3.9e-3} & \textbf{98.72} & 1.2e-3 & \textbf{89.60} & \textbf{1.7e-2} & 62.38 & \textbf{8.5e-1} & \textbf{61.58} & \textbf{1.1e0} & \textbf{62.90} & \textbf{1.5e0} \\
    \bottomrule
    \end{tabular}
    }
\end{table*}

\begin{table*}[!t]
  \centering
      \caption{\method and baselines: results of test accuracy (\%) and fairness ($\Psi$) in different heterogeneity and scalability, while the privacy budget $\epsilon=3.52$.
      }
    \label{tab:exp2-2}
  \resizebox{1.0\linewidth}{!}{
    \begin{tabular}{l|*{6}{c}|*{6}{c}}
    \toprule
    & \multicolumn{6}{c|}{Heterogeneity} & \multicolumn{6}{c}{Scalability}\\
    \midrule
    Datasets & \multicolumn{4}{c}{MNIST} & \multicolumn{2}{c|}{FashionMNIST} & \multicolumn{6}{c}{CIFAR10} \\
    \midrule
    Method & \multicolumn{2}{c}{$\mathrm{Dir}(0.1)$} & \multicolumn{2}{c}{$\mathrm{Dir}(0.5)$} & \multicolumn{2}{c|}{$\mathrm{Dir}(1)$}
    & \multicolumn{2}{c}{10 clients} & \multicolumn{2}{c}{20 clients} & \multicolumn{2}{c}{50 clients}\\
    \midrule
     & Acc.($\uparrow$) & $\Psi(\downarrow)$
     & Acc.($\uparrow$) & $\Psi(\downarrow)$
     & Acc.($\uparrow$) & $\Psi(\downarrow)$
     & Acc.($\uparrow$) & $\Psi(\downarrow)$
     & Acc.($\uparrow$) & $\Psi(\downarrow)$
     & Acc.($\uparrow$) & $\Psi(\downarrow)$
     \\
    \midrule
    FedAvg & 93.40 & 1.1e11 & 94.17 & 4.1e9 & 85.60 & 3.7e9 &52.61 & 6.1e9 & 51.58 & 2.0e9 & 55.90 & 1.5e9 \\
    SCAFFOLD & 93.95 & 7.0e10 & 94.06 & 1.9e9 & 84.56 & 3.6e9 &53.85 & 6.2e9 & 54.34 & 2.9e8 & 53.98 & 2.2e9 \\
    FedProx & 90.50 & 5.3e10 & 94.23 & 3.1e9 & 84.38 & 5.6e9 &51.34 & 4.4e9 & 53.67 & 4.5e8 & 54.14 & 2.2e9 \\
    FedDyn & 91.42 & 5.6e10 & 93.08 & 3.8e9 & 85.21 & 9.8e10 &52.14 & 5.5e9 & 50.21 & 1.2e10 & 51.86 & 2.0e10 \\
    ALI-DPFL & 90.89 & 3.3e10 & 93.98 & 2.6e9 & 83.66 & 2.5e10 &52.05 & 3.6e9 & 38.35 & 2.4e10 & 33.46 & 1.6e10 \\
    q-FFL & 93.74 & 7.8e10 & 93.86 & \textbf{2.5e8} & 85.20 & 7.4e8 &48.46 & 4.7e9 & 51.12 & 2.7e8 & 49.83 & \textbf{1.1e7} \\
    \method & \textbf{94.13} & \textbf{2.3e10} & \textbf{94.56} & 3.4e9 & \textbf{86.77} & \textbf{6.0e8} &\textbf{54.21} & \textbf{2.6e9} & \textbf{54.52} & \textbf{2.2e8} & \textbf{56.49} & 5.3e8 \\
    \bottomrule
    \end{tabular}
    }
\end{table*}

\textbf{Heterogeneity.} Table \ref{tab:exp2-1} shows that \premethod outperforms baseline methods in accuracy while maintaining significant fairness when faced with diverse data distributions exhibiting varying levels of heterogeneity. Except for the $\mathrm{Dir}(0.5)$ scenario, where it falls slightly short of q-FFL, \premethod demonstrated fairness improvements of 22.0\% and 50\% over the best baseline in $\mathrm{Dir}(0.1)$ and $\mathrm{Dir}(1.0)$, respectively. Table \ref{tab:exp2-2} shows that \method outperforms baseline methods in accuracy while maintaining significant fairness when confronted with diverse data distributions exhibiting varying levels of heterogeneity. Except for the $\mathrm{Dir}(0.5)$ scenario, where it slightly underperforms q-FFL, \method shows fairness improvements of 30.3\% and 18.9\% over the best baseline in $\mathrm{Dir}(0.1)$ and $\mathrm{Dir}(1.0)$, respectively.

\textbf{Scalability.} Table \ref{tab:exp2-1} indicates that \premethod generally surpasses baselines in accuracy and shows fairness improvements of 9.6\%, 15.4\%, and 0.01\% across three client counts. Table \ref{tab:exp2-2} demonstrates that as the number of clients increases, \method consistently achieves accuracy comparable to the best baseline. Moreover, except in the 50 clients scenario where it slightly lags behind q-FFL in fairness, \method exhibits fairness improvements of 27.8\% and 18.5\% with 10 and 20 clients, respectively.

\begin{figure}[!t]
	\begin{center}      
		\includegraphics[width=0.8\linewidth]{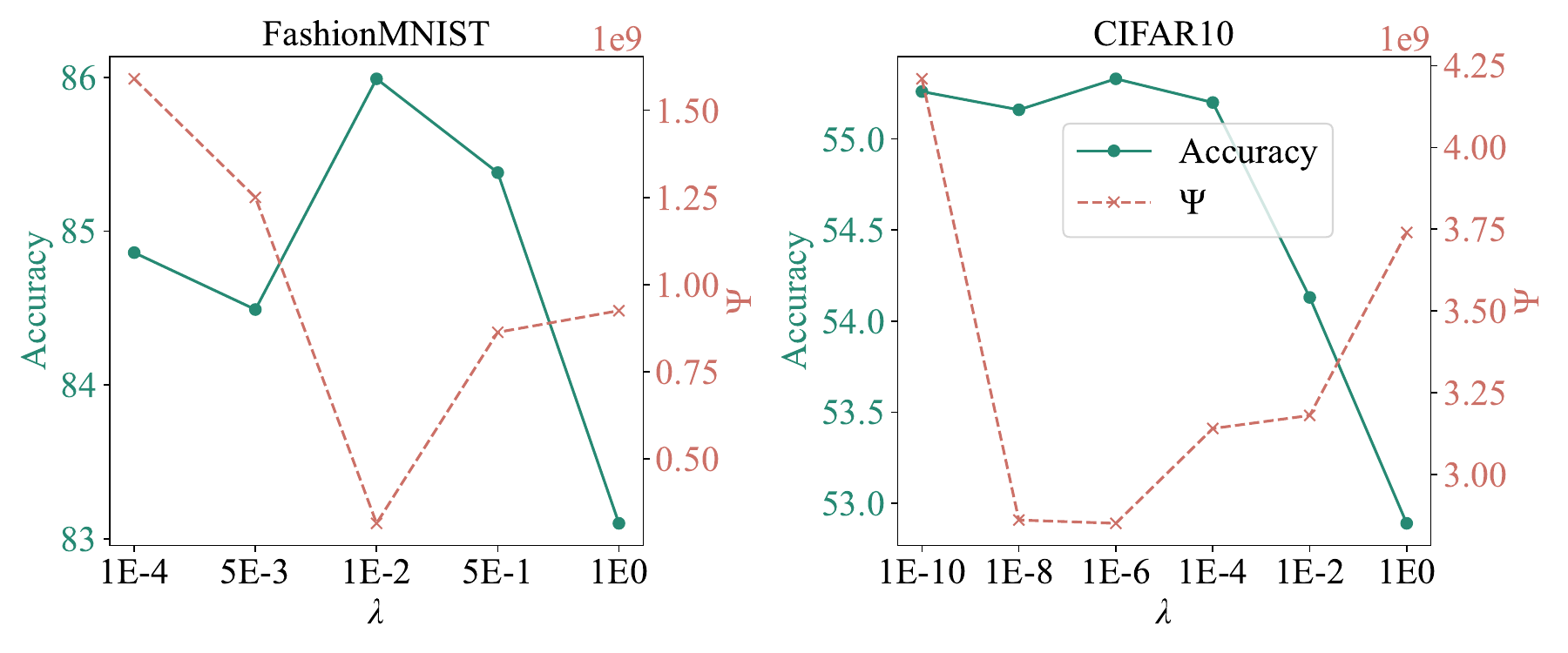}
        \vspace{-0.5cm}
		\caption{Impact of different $\lambda$ in \method algorithm.}
		\label{fig:optimal_lambda}
	\end{center}
\end{figure}

\begin{figure}[!t] 
	\begin{center}
		\includegraphics[width=0.8\linewidth]{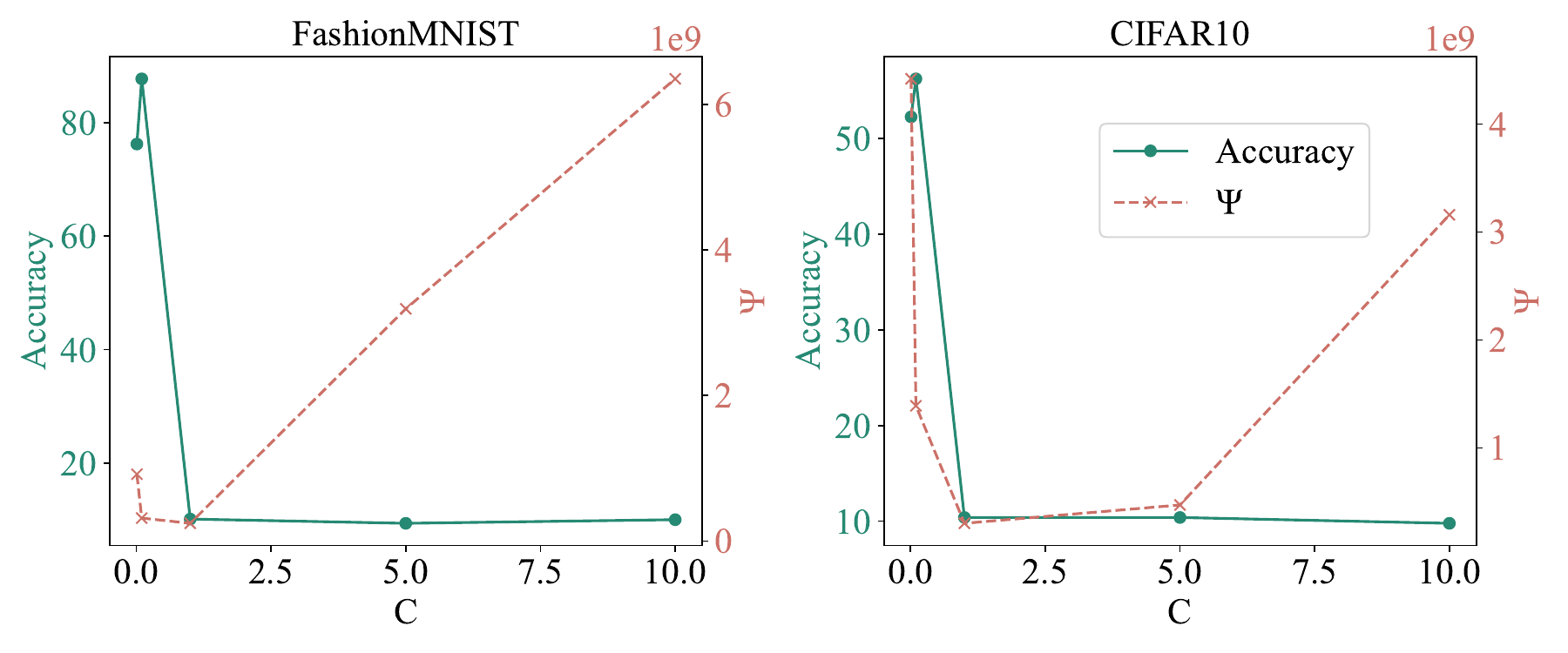}
        \vspace{-0.5cm}
		\caption{Impact of different $C$ in \method algorithm.}
        \label{fig:optimal_C}
	\end{center}
\end{figure}

\begin{figure}[!t] 
	\begin{center}
		\includegraphics[width=0.8\linewidth]{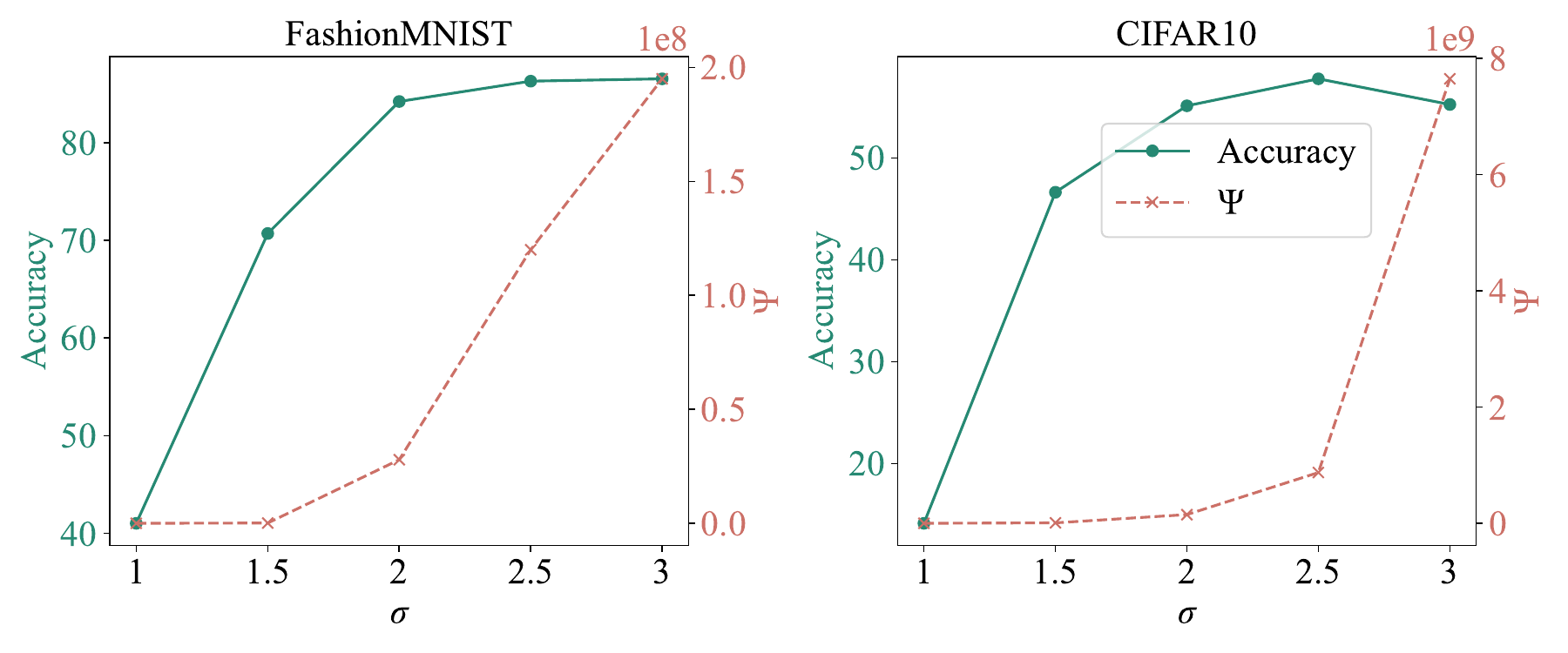}
        \vspace{-0.5cm}
		\caption{Impact of different $\sigma$ in \method algorithm. }
        \label{fig:sigma_acc_psi}
	\end{center}
\end{figure}

\subsection{Performance under Different Hyper-parameters}\label{subsec:exp_3}
In response to \textbf{RQ3}, we performed extensive hyper-parameter experiments. In Table~\ref{table:diff_eps_experiments}, we observed results under fixed privacy budgets $\epsilon = \{1.0, 2.0, 3.0, 4.0\}$. Table~\ref{table:diff_target_accuracy_experiments} presents experiments aiming at fixed target accuracies $\{80\%, 85\%, 90\%, 95\%\}$. Fig.~\ref{fig:optimal_lambda} explores the existence of an optimal $\lambda$, with results consistent with those in Section~\ref{sec:guarantees}. The influence of the clipping norm $C$ is investigated in Fig.~\ref{fig:optimal_C}, and the effect of the noise multiplier $\sigma$ is examined in Fig.~\ref{fig:sigma_acc_psi}.

\textbf{Impact of Different $\lambda$.} In Section \ref{subsec:optimal_lambda}, we derive a closed-form solution for $\lambda^*$, which, however, relies on constants or upper bounds (e.g., $L$, $\mu$, $G$, $\Gamma$) whose specific values change with different models, datasets, and loss functions. Therefore, we conduct the experiment in Fig. \ref{fig:optimal_lambda} to investigate the practical range of $\lambda^*$. We investigated the correlation between $\lambda$ and $\Psi$ in the \method algorithm. As depicted in Fig. \ref{fig:optimal_lambda}, our findings align with the theoretical analysis: an initial increase in $\lambda$ leads to a decrease in $\Psi$, followed by an increase, indicating the existence of an optimal $\lambda$.

\textbf{Impact of Different $C$.} We emphasized that excessively large or small values of $C$ cannot ensure both performance and fairness simultaneously. As illustrated in Fig. \ref{fig:optimal_C}, we examined the correlation between $C$, performance, and fairness across two datasets, with $C$ taking values in the range \{0.01, 0.1, 1, 5, 10\}. Although $\Psi$ reaches its minimum when $C=1$, the model does not converge due to excessive noise addition. As $C$ continues to increase, the model's performance further deteriorates, resulting in an increase in $\Psi$. When $C=0.1$, both accuracy and $\Psi$ achieve relatively good performance.

\textbf{Impact of Different $\sigma$.} We tested the variations in accuracy and fairness on two datasets when $\sigma$ was set to $\{1, 1.5, 2, 2.5, 3\}$, under the condition of $\epsilon=2$ and $q=0.05$, which supports $T \in \{6, 115, 268, 463, 708\}$ rounds of communication. As shown in Fig. \ref{fig:sigma_acc_psi}, an increase in $\sigma$ allows more training epochs to advance the model's convergence process, but when $\sigma=3$, excessive noise hinders the convergence. When $\sigma$ is smaller, the model performance is relatively similar, with a lower $\Psi$ value. As $\sigma$ gradually increases, the performance differences between models become apparent, and the $\Psi$ value increases.

\begin{table*}[!t]  
\centering  
\caption{The average accuracy ($\%$) and fairness ($\Psi$) on different privacy budget ($\epsilon$) setting at FashionMNIST dataset.}  
\label{table:diff_eps_experiments}  
\resizebox{0.8\linewidth}{!}{
\begin{tabular}{l|*{2}{c}|*{2}{c}|*{2}{c}|*{2}{c}}
\toprule
\textbf{$\epsilon$} & \multicolumn{2}{c|}{1.0} & \multicolumn{2}{c|}{2.0}& \multicolumn{2}{c|}{3.0}& \multicolumn{2}{c}{4.0} \\  
\midrule
& Acc.($\uparrow$) & $\Psi(\downarrow)$& Acc.($\uparrow$) & $\Psi(\downarrow)$& Acc.($\uparrow$) & $\Psi(\downarrow)$& Acc.($\uparrow$) & $\Psi(\downarrow)$ \\

\midrule
FedAvg & 61.68 & 1.1e6 & 82.69 & 2.3e7 & 84.07 & 3.8e8 & 86.83 & 8.0e8 \\  
SCAFFOLD & 62.12 & 1.2e6 & 82.36 & 2.3e7 & 83.39 & 2.6e9 & 85.41 & 5.6e8 \\  
FedProx & 61.25 & 1.1e6 & 81.62 & 2.3e7 & 85.32 & 3.8e8 & 85.68 & 2.4e9 \\  
FedDyn & 62.38 & 1.5e6 & 82.43 & 3.1e7 & 83.29 & 2.9e8 & 84.51 & 5.2e9 \\  
ALI-DPFL & 61.78 & 1.1e6 & 83.01 & 2.3e7 & 84.12 & 3.9e8 & 85.58 & 4.5e9 \\  
q-FFL & 58.99 & 3.6e6 & 80.24 & 9.6e7 & 82.92 & 5.5e8 & 85.01 & 3.8e9 \\  
\method & \textbf{63.36} & \textbf{1.0e6} & \textbf{83.15} & \textbf{2.1e7} & \textbf{85.59} & \textbf{9.5e7} & \textbf{86.97} & \textbf{3.8e8} \\  
\bottomrule
\end{tabular}  
}

\end{table*}  

\textbf{Impact of Different $\epsilon$.} We conducted experiments with $\epsilon \in \{1, 2, 3, 4\}$ on the FashionMNIST dataset, as detailed in Table \ref{table:diff_eps_experiments}. For $\epsilon \in \{1, 2, 3\}$, we allowed larger values of $\epsilon$ to support more iterations. At $\epsilon = 4$, we supported a smaller noise multiplier $\sigma = 1.65$ with a similar number of iterations as at $\epsilon = 3$. The experimental results demonstrate that \method consistently achieves performance on par with the best baseline in terms of accuracy across various privacy budgets. In terms of fairness, the improvements are 9.1\%, 8.7\%, 67.2\%, and 32.1\%, respectively.

\begin{table*}[!t]  
\centering  
\caption{The privacy budget ($\epsilon$) and fairness ($\Psi$) on different target accuracy setting at MNIST $\mathrm{Dir}(0.05)$.}
\label{table:diff_target_accuracy_experiments}  
\resizebox{0.8\linewidth}{!}{
\begin{tabular}{l|*{2}{c}|*{2}{c}|*{2}{c}|*{2}{c}}
\toprule
\textbf{Acc.} & \multicolumn{2}{c|}{80\%} & \multicolumn{2}{c|}{85\%}& \multicolumn{2}{c|}{90\%}& \multicolumn{2}{c}{95\%} \\  
\midrule
& $\epsilon(\downarrow)$ & $\Psi(\downarrow)$& $\epsilon(\downarrow)$ & $\Psi(\downarrow)$& $\epsilon(\downarrow)$ & $\Psi(\downarrow)$& $\epsilon(\downarrow)$ & $\Psi(\downarrow)$ \\  
\midrule
FedAvg & 2.43 & 2.10e9 & 3.02 & 8.90e9 & 3.45 & 1.30e10 & 4.89 & 1.30e11 \\  
SCAFFOLD & 2.55 & 3.30e9 & 3.15 & 1.10e10 & 3.67 & 3.70e10 & 5.32 & 1.10e11 \\  
FedProx & 3.02 & 1.30e9 & 2.98 & 1.30e10 & 4.31 & 5.80e10 & 4.89 & 2.50e11 \\  
FedDyn & 2.68 & 2.20e9 & 3.55 & 7.80e9 & 3.98 & 1.10e10 & 5.15 & 3.30e11 \\  
ALI-DPFL & 2.85 & 3.10e9 & 2.98 & 1.30e10 & 3.67 & 2.30e10 & 5.32 & 9.80e10 \\  
q-FFL & 2.68 & 9.80e8 & 3.15 & 9.70e9 & 4.12 & 8.80e10 & 4.04 & 1.02e11 \\  
\method & \textbf{2.25} & \textbf{6.60e8} & \textbf{2.70} & \textbf{1.20e9} & \textbf{3.05} & \textbf{5.80e9} & \textbf{3.77} & \textbf{2.80e10} \\  
\bottomrule
\end{tabular}  
}  
\end{table*}

\textbf{$\epsilon$ at Different Target Accuracy.} TABLE \ref{table:diff_target_accuracy_experiments} presents experiments on MNIST under the $\mathbf{Dir}(0.05)$ partition, reporting the privacy budget $\epsilon$ and fairness measure $\Psi$ at the point of achieving accuracy targets $\{80\%,85\%,90\%,95\%\}$. As the target accuracy increases (from $80\%$ to $95\%$), the privacy budget ($\epsilon$) for most methods rises slightly (e.g., FedAvg increases from 2.43 to 3.02), while fairness ($\Psi$) deteriorates significantly (e.g., FedAvg rises from 2.10×10$^9$ to 1.3×10$^{11}$), highlighting the intensified trade-off between privacy and fairness under higher accuracy demands. FedFDP demonstrates exceptional performance in balancing these objectives: At the high $95\%$ accuracy target, its $\epsilon$ (2.70) is lower than all other methods, and its $\Psi$ (1.20×10$^{9}$) is significantly lower than competitors by an order of magnitude. At the $80\%$ accuracy target, FedFDP achieves the lowest $\Psi$ (6.60×10$^{8}$) and the lowest $\epsilon$ (2.25) in the entire table, validating its comprehensive superiority.

\begin{figure}[!t] 
	\begin{center}
		\includegraphics[width=0.8\linewidth]{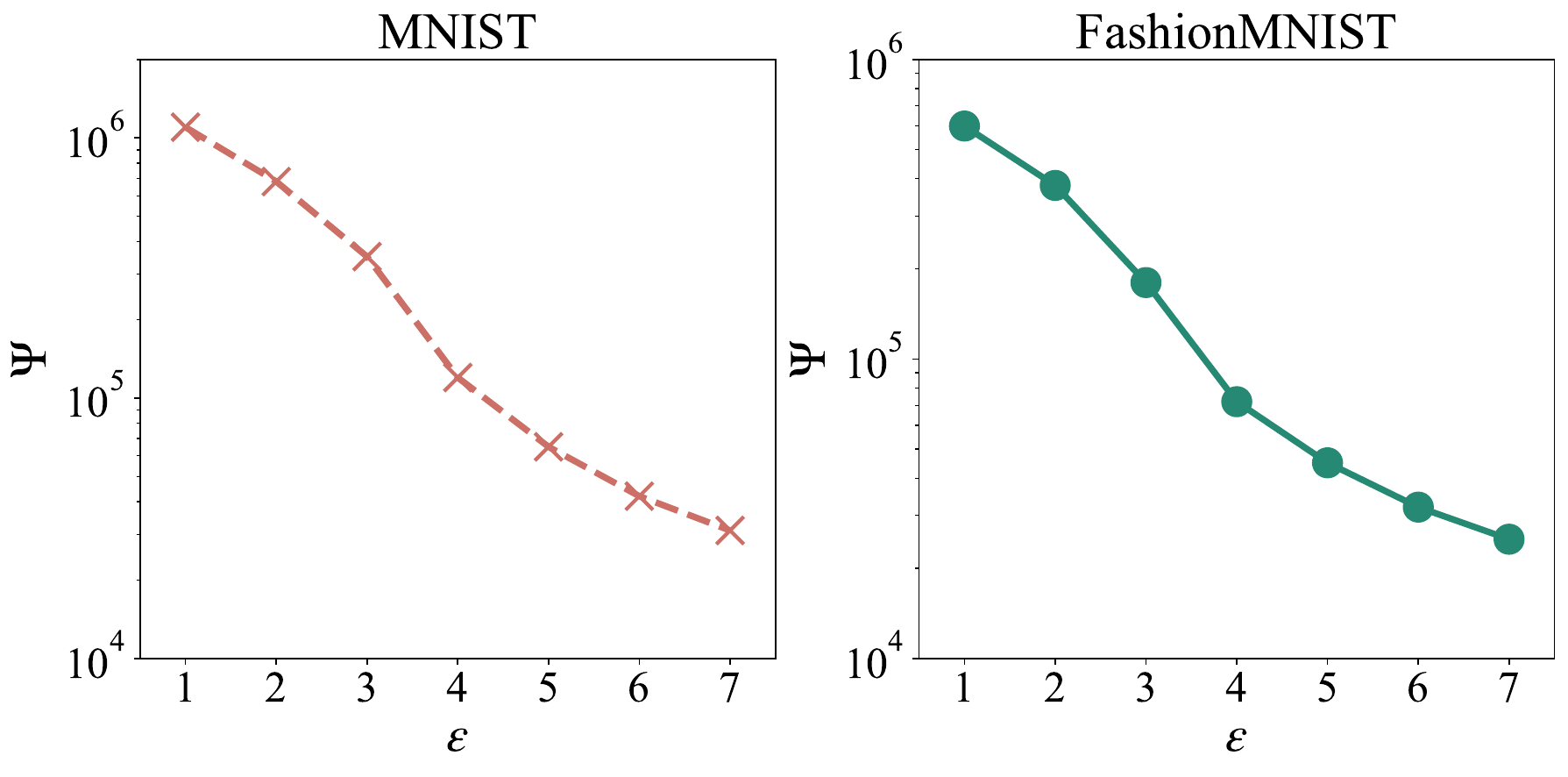}
        \vspace{-0.5cm}
		\caption{Tradeoff between privacy and fairness on $60\%$ accuracy.}
        \label{fig:psi_vs_epsilon}
	\end{center}
\end{figure}

\textbf{Tradeoff between Privacy and Fairness.} With the target accuracy fixed at $60\%$ on both MNIST and FashionMNIST, we investigate the trade-off between fairness and privacy. As shown in Fig. \ref{fig:psi_vs_epsilon}, as the privacy budget $\epsilon$ increases, the fairness metric $\Psi$ decreases significantly on both datasets, indicating a clear trade-off between privacy and fairness under the given accuracy constraint. The noise introduced by the privacy mechanism, while protecting data, amplifies performance disparities across groups, whereas relaxing privacy constraints creates room for fairness optimization. This monotonically decreasing relationship reveals the tripartite trade-off characteristics among privacy, fairness, and accuracy.

\vspace{-0.2cm}
\section{Limitations and Future Work}

Despite the promising results achieved by FedFDP in balancing fairness, privacy, and utility, there are several limitations in the current study that open avenues for future research.

First, regarding heterogeneity, our current evaluation primarily focuses on statistical heterogeneity (Non-IID data) simulated via Dirichlet partitioning. We have not yet deeply explored \textit{model heterogeneity}, where clients may possess different network architectures due to varying storage or memory capacities. Furthermore, the challenges associated with \textit{cross-domain} data—where feature spaces or distributions differ fundamentally across clients—remain unaddressed. Future work could investigate adapting the fairness-aware gradient clipping strategy to support heterogeneous model architectures and cross-domain federated learning scenarios.

Second, the current framework assumes that all selected clients participate successfully in the training process (full participation). However, in practical real-world deployments, \textit{device heterogeneity} (e.g., varying computational power) and unstable network conditions (e.g., bandwidth limits and high latency) often lead to the problem of \textit{stragglers}. These stragglers can significantly delay the synchronization process in the global aggregation phase. In this work, we have not specifically optimized FedFDP for such system-level challenges. Future iterations of FedFDP could explore asynchronous update mechanisms or robust client selection protocols to mitigate the impact of stragglers and communication dropouts.
\vspace{-0.2cm}
\section{Conclusion}\label{sec:conclusion}

In this paper, a \premethod algorithm is initially proposed to effectively address fairness concerns by optimizing a novel local loss function. The \method, building upon \premethod, is introduced to incorporate a fairness-aware gradient clipping strategy and an adaptive clipping method for additional loss values, thereby achieving both fairness and differential privacy protection. Then, we find an optimal fairness parameter $\lambda^*$ through convergence analysis and numerical analysis methods, striking a balance between model performance and fairness. Subsequently, a comprehensive privacy analysis of the approach is conducted using RDP. Through extensive experiments, the results indicate that \premethod and \method significantly outperform state-of-the-art solutions in terms of model performance and fairness. It is believed that our work contributes a valuable FL framework for addressing fairness and privacy challenges.

\section*{Acknowledge}
This work is supported by National Natural Science Foundation of China Key
Program (Grant No. 62132005), Natural Science Foundation of Shanghai (Grant No.
22ZR1419100), and CAAI-Huawei MindSpore Open
Fund (Grant No. CAAIXSJLJJ-2022-005A).


\bibliographystyle{splncs04}
\bibliography{cite}

\clearpage

\onecolumn
\appendix
\section{Appendix}\label{sec:appendix}

In the beginning, Table \ref{table:notations} presents some notations utilized in this paper. Symbols that have not been previously introduced will be defined in subsequent sections.
\begin{table}[tbh]
\centering
\caption{Summary of main notations}
\label{table:notations}
\begin{tabular}{p{0.1\textwidth} p{0.80\textwidth}} 
\hline
$F(\mathbf{w}$) & Global loss function  \hfill\\
$F_i(\mathbf{w}$) & Local loss function for client $i$  \hfill\\
$T$ & Communication rounds\hfill\\
$t$ & Communication round index  \hfill\\
$D_i$ & Dataset of client $i$ \hfill \\
$D$ & The union of $D_i$ \hfill \\
$\mathbf{w}_t^i$ & Local model at client $i$ at round $t$  \hfill\\
$\mathbf{w}^*$ & Optimal model that minimizes $F(\mathbf{\mathbf{w}})$ \hfill\\
$\eta$ & Gradient descent step size \hfill\\
$\lambda$ & Fairness parameter\hfill\\
$N$ & Total number of clients \hfill\\
$q$ & batch sample ratio \hfill\\
$|\mathcal{B}_i|$ & Batch size of client $i$, with the mathematical expectation equals to $|D_i|\cdot q$ \hfill \\
$p_i$ & Weight of client $i$, equals to $|D_i|/|D|$\hfill \\
$\xi_j$ & Single data sample in $D_i$, indexed by $j$ \hfill\\
$\sigma$ & Noise multiplier for gradient\hfill \\
$\sigma_l$ & Noise multiplier for loss\hfill \\
$C$ & Original clipping bound for gradient\hfill \\
$C_l$ & Original clipping bound for loss\hfill \\
$||\cdot||$ & $L_2$-norm \hfill \\
\hline
\end{tabular}
\end{table}

\subsection{Convergence Analysis}\label{subsec:ca}
We analyze the convergence of Algorithm \ref{alg:method} and derive insights on selecting the hyperparameter $\lambda$ based on the convergence upper bound. Prior to proving this, we need to establish several assumptions:

\begin{assumption}\label{ass:l_smooth}
    $F_1, \cdots, F_N$ are all L-smooth: for all $\mathbf{v}$ and $\mathbf{w}, F_i(\mathbf{v}) \leq F_i(\mathbf{w})+(\mathbf{v}-\mathbf{w})^\top \nabla F_i(\mathbf{w})+\frac{L}{2}\|\mathbf{v}-\mathbf{w}\|^2$.
\end{assumption}

\begin{assumption}\label{ass:mu_strong_convex}
    $F_1, \cdots, F_N$ are all $\mu$-strongly convex: for all $\mathbf{v}$ and $\mathbf{w}, F_i(\mathbf{v}) \geq F_i(\mathbf{w})+(\mathbf{v}-\mathbf{w})^\top \nabla F_i(\mathbf{w})+\frac{\mu}{2}\|\mathbf{v}-\mathbf{w}\|^2$.
\end{assumption}

\begin{assumption}\label{ass:grad_bound}
    Let $\xi_j$ be sampled from the $i$-th device's local data uniformly at random. The expected squared norm of stochastic gradients is uniformly bounded, i.e., $\left\|\nabla F_i\left(\mathbf{w}_t^i, \xi_j\right)\right\| \leq G$ for all $i\in[1, \cdots, N]$ , $t\in[0, \cdots, T-1]$  and $j\in[1,\cdots,|\mathcal{B}_i|]$
\end{assumption}



\begin{theorem}\label{theo:ca}
    Under the assumptions mentioned above, we obtain the convergence upper bound for Algorithm \ref{alg:method} as follows:
\end{theorem}

\begin{equation}\label{eq:ca}
    \mathbb{E}[F(\mathbf{w}_t)] - F^* \leq \frac{L}{2t}\left( \frac{A}{\mu^2 (2C_t -1)} + \mathbb{E}\|\mathbf{w}_1 - \mathbf{w}^*\|^2 \right),
\end{equation}

where: 
\begin{itemize}
    \item $A = G^2 C_t^3 + 3 G^2 C_t^2 + 2 L \Gamma C_t + \frac{2\sigma^2 C^2 d}{\hat{B}^2}$,
    \item $\hat B = \min_i |\mathcal{B}_i|$, 
    \item $\Gamma = F^* - \sum_{i=1}^{N} p_i F_i^* $,
    \item $C_t = \sum_{i=1}^{N} p_i C_t^i, C_t^i = \frac{1}{|\mathcal{B}_i|}\sum_{j=1}^{|\mathcal{B}_i|} C_t^{i,j} $.
\end{itemize}
The meaning of $\Gamma$ is consistent with \cite{XiangLi2020OnTheConvergence}, where a larger value of $\Gamma$ indicates that the data among different clients is more Non-IID.

\textit{proof sketch:} We investigate the relationship between $\mathbb{E}\| \mathbf{w}_{t+1} - \mathbf{w}^{*} \|^2$ and $\mathbb{E}\| \mathbf{w}_{t} - \mathbf{w}^{*} \|^2$, and then use mathematical induction to obtain an upper bound for $\mathbb{E}\| \mathbf{w}_{t} - \mathbf{w}^{*} \|^2$. Finally, by utilizing Assumption \ref{ass:l_smooth}, we derive Equation (\ref{eq:ca}). For the detailed proof, please refer to Appendix \ref{append:proof_of_ca}.


\subsection{Proof of Theorem \ref{theo:ca}}\label{append:proof_of_ca}
For the purpose of validation, we introduce an additional variable $\mathbf{v}_t^i$ to represent the immediate result of a single-step DPSGD update from $\mathbf{w}_t^i$. We interpret $\mathbf{w}_{t+1}^i$ as the parameter obtained after a single communication step. Consequently, the fair-clipping DPSGD in client $i$ at iteration $t$ transitions from Equation (\ref{eq:sgd_dynamic_clipping}) to:

\begin{equation}
    \mathbf{v}_{t+1}^i
    = \mathbf{w}_t^{i} - \frac{\eta}{|\mathcal{B}_i|}[
    \sum^{|\mathcal{B}_i|}_{j=1} C^{i,j}_t \cdot \nabla F_i(\mathbf{w}_t^i,\xi_j)
    + \sigma C \cdot \mathcal{N}(0,\mathbf{I})],
\end{equation}
where:
\begin{equation*}
    C^{i,j}_t = \min\left( 1+\lambda\cdot\Delta_{i}^j,\frac{C}{\| \nabla F_i(\mathbf{w}_t^i,\xi_j) \|} \right).
\end{equation*}
In our analysis, we define two virtual sequences $\mathbf{v}_t=\sum_{i=1}^N p_i \mathbf{v}_t^i$ and $\mathbf{w}_t=\sum_{i=1}^N p_i \mathbf{w}_t^i$, which is motivated by \cite{stich2018local}. Therefore,
\begin{align}\label{eq:barv_barw_gd}
    \mathbf{v}_{t+1}
    = \mathbf{w}_t 
    - \sum_{i=1}^{N}p_i \frac{\eta}{|\mathcal{B}_i|}[
    \sum^{|\mathcal{B}_i|}_{j=1} C^{i,j}_t \cdot \nabla F_i(\mathbf{w}_t^i,\xi_j)
    + \sigma C \cdot \mathcal{N}(0,\mathbf{I})]
\end{align}

\subsubsection{Key Lemma}

\begin{lemma}\label{lem:result_one_iteration}
    (Results of one iteration.)Assume Assumption \ref{ass:l_smooth}-\ref{ass:grad_bound} hold, we have:
\begin{equation}
    \mathbb{E}\| \mathbf{v}_{t+1} - \mathbf{w}^* \|^2 \nonumber\leq  \left( 1-\mu\eta C_t \right) 
    \mathbb{E}\| \mathbf{w}_{t} - \mathbf{w}^* \|^2
    + \eta^2 A,
\end{equation}
\end{lemma}

where: 
\begin{itemize}
    \item $A = G^2 C_t^3 + 3 G^2 C_t^2 + 2 L \Gamma C_t + \frac{2\sigma^2 C^2 d}{\hat{B}^2} $,
    \item $\hat B = \min_i |\mathcal{B}_i|$,
    \item $\Gamma = F^* - \sum_{i=1}^{N} p_i F_i^* $,
    \item $C_t = \sum_{i=1}^{N} p_i C_t^i, C_t^i = \frac{1}{|\mathcal{B}_i|}\sum_{j=1}^{|\mathcal{B}_i|} C_t^{i,j} $.
\end{itemize}

Let $\Delta_t = \mathbb{E}\| \mathbf{w}_{t} - \mathbf{w}^*\|^2$. It is evident that we always have $\mathbf{w}_{t+1}=\mathbf{v}_{t+1}$. According to Lemma \ref{lem:result_one_iteration}, this implies:
\begin{equation*}
    \Delta_{t+1} \leq \left( 1-\mu\eta C_t \right)  \Delta_{t} + \eta^2 A
\end{equation*}

We use mathematical induction to obtain $\Delta_t \leq \frac{v}{t}$ where $v=\max \{ \frac{\beta^2 A}{\mu\beta C_t -1}, \Delta_1 \}$, $\eta=\frac{\beta}{t}$ for some $\beta>\frac{1}{\mu}$.

\textbf{STEP 1.} When $t=1$, the equation $\Delta_1 \leq v$ holds obviously.

\textbf{STEP 2.} We assume $\Delta_t \leq \frac{v}{t}$ holds.

\textbf{STEP 3.} 
\begin{align*}
    \Delta_{t+1} 
    \leq \left(1- \mu\frac{\beta}{t}C_t \right) \frac{v}{t} + \frac{\beta^2 A}{t^2} 
    = \frac{t-1}{t^2}v + \left( \frac{\beta^2 A}{t^2} - \frac{\mu\beta C_t -1}{t^2}v \right) 
    \leq \frac{t-1}{t^2}v 
    \leq \frac{v}{t+1} 
\end{align*}

Therefore, $\Delta_{t+1} \leq \frac{v}{t+1}$ holds, completing the proof by mathematical induction. Hence, $\Delta_t \leq \frac{v}{t}$ holds.

Then by the $L$-smoothness of $F(\cdot)$, let $\beta=\frac{2}{\mu}$, we get

\begin{equation*}
    \mathbb{E}[F(\mathbf{w}_t)]-F^* \leq \frac{L}{2}\Delta_t 
    \leq \frac{L}{2t}v 
    \leq \frac{L}{2t}\left( \frac{A}{\mu^2(2 C_t -1)} + \Delta_1 \right)
\end{equation*}

\subsubsection{Proof of Lemma \ref{lem:result_one_iteration}}
By the Equation (\ref{eq:barv_barw_gd}), we get
\begin{align*}
&\left\| \mathbf{v}_{t+1}-\mathbf{w}^{*} \right\|^2\\
&= \| \mathbf{w}_{t} -\mathbf{w}^{*} 
- \sum_{i=1}^{N}p_i \frac{\eta}{|\mathcal{B}_i|}[
    \sum^{|\mathcal{B}_i|}_{j=1} C^{i,j}_t \cdot \nabla F_i(\mathbf{w}_t^i,\xi_j)
    + \sigma C \cdot \mathcal{N}(0,\mathbf{I})]
\|^2\\
&= \left\| \mathbf{w}_{t}-\mathbf{w}^{*} \right\|^2\\
&\underbrace{- 2\langle \mathbf{w}_{t}-\mathbf{w}^{*}, 
    \sum_{i=1}^{N}p_i \frac{\eta}{|\mathcal{B}_i|}[
    \sum^{|\mathcal{B}_i|}_{j=1} C^{i,j}_t  \nabla F_i(\mathbf{w}_t^i,\xi_j)
    + \sigma C \mathcal{N}(0,\mathbf{I})]}_{\mathcal{A}_1}
\rangle \\
&+ \underbrace{ 
    \left\|\sum_{i=1}^{N}p_i \frac{\eta}{|\mathcal{B}_i|}[
    \sum^{|\mathcal{B}_i|}_{j=1} C^{i,j}_t \cdot \nabla F_i(\mathbf{w}_t^i,\xi_j)
    + \sigma C \cdot \mathcal{N}(0,\mathbf{I})]\right\|^2
}_{\mathcal{A}_2}
\end{align*}

Firstly, we process $\mathcal{A}_2$:

\begin{align*}
    &\mathcal{A}_2 \leq 
    \underbrace{\left\|\sum_{i=1}^{N}p_i \frac{\eta}{|\mathcal{B}_i|}
    \sum^{|\mathcal{B}_i|}_{j=1} C^{i,j}_t \cdot \nabla F_i(\mathbf{w}_t^i,\xi_j) \right\|^2}_{\mathcal{B}_1}
     \\
    &\quad + 
    \underbrace{\sum_{i=1}^{N}p_i \frac{\eta}{|\mathcal{B}_i|}\left\langle
    \sum^{|\mathcal{B}_i|}_{j=1} C^{i,j}_t \cdot \nabla F_i(\mathbf{w}_t^i,\xi_j), \sigma C \cdot \mathcal{N}(0,\mathbf{I})\right\rangle}_{\mathcal{B}_0}
     \\
    &\quad +
    \underbrace{\| \sum_{i=1}^{N}p_i \frac{\eta}{|\mathcal{B}_i|} \sigma C \cdot \mathcal{N}(0,\mathbf{I})\|^2}_{\mathcal{B}_2}
    \\
\end{align*}

Since $\mathbb{E}[\mathcal{B}_0] = 0$, we focus on $\mathcal{B}_1$ and $\mathcal{B}_2$:

\begin{align*}
    \mathbb{E}[\mathcal{B}_2] \leq \frac{\eta^2}{\hat{B}^2} \sum_{i=1}^N
    \mathbb{E} \| \sigma C \mathcal{N}(0,\mathbf{I}) \|^2
    \leq \frac{\eta^2 \sigma^2 C^2 d}{\hat{B}^2},
\end{align*}

where $\frac{1}{\hat{B}^2} = \max_i \frac{1}{|\mathcal{B}_i|}$.

By the convexity of $\|\cdot\|^2$,
\begin{align*}
    \mathcal{B}_1 \leq  \eta^2 \sum_{i=1}^{N}p_i \left\| \frac{1}{|\mathcal{B}_i|}
    \sum^{|\mathcal{B}_i|}_{j=1} C^{i,j}_t \cdot \nabla F_i(\mathbf{w}_t^i,\xi_j) \right\|^2,
\end{align*}
taking exception and according to assumption \ref{ass:grad_bound}:

\begin{align*}
    \mathbb{E}[\mathcal{B}_1] \leq \eta^2 \mathbb{E} \left[\sum_{i=1}^{N}p_i \left\| C_t^i \nabla F_i(\mathbf{w}_t^i) \right\|^2\right]
    \leq \eta^2 C_t^2 G^2,
\end{align*}
where $C_t=\sum_{i=1}^N p_i C_t^i$, $C_t^i=\frac{1}{|\mathcal{B}_i|}\sum_{j=1}^{|\mathcal{B}_i|} C_t^{i,j}$.

Now, we obtain the bound for the expectation of $\mathcal{A}_2$:
\begin{align*}
    \mathbb{E}[\mathcal{A}_2] &\leq \eta^2 
    \left( \frac{\sigma^2 C^2 d}{\hat{B}^2} + G^2 C_t^2 \right)
\end{align*}

The process of $\mathcal{A}_1$ show as below:
\begin{align*}
    \mathcal{A}_1 &= 
    - 2\langle \mathbf{w}_{t}-\mathbf{w}^{*}, 
    \sum_{i=1}^{N}p_i \frac{\eta}{|\mathcal{B}_i|}
    \sum^{|\mathcal{B}_i|}_{j=1} C^{i,j}_t  \nabla F_i(\mathbf{w}_t^i,\xi_j) \rangle
    \underbrace{- 2\langle \mathbf{w}_{t}-\mathbf{w}^{*}, 
    \sum_{i=1}^{N}p_i \frac{\eta}{|\mathcal{B}_i|} \sigma C \mathcal{N}(0,\mathbf{I})\rangle}_{\mathcal{C}_0}
    \\&= \mathcal{C}_0 
    \underbrace{- 2\sum_{i=1}^{N}p_i\langle \mathbf{w}_{t}-\mathbf{w}_{t}^{i}, 
     \frac{\eta}{|\mathcal{B}_i|}
    \sum^{|\mathcal{B}_i|}_{j=1} C^{i,j}_t  \nabla F_i(\mathbf{w}_t^i,\xi_j) \rangle}_{\mathcal{C}_1}
    \underbrace{- 2\sum_{i=1}^{N}p_i\langle \mathbf{w}_{t}^{i} - \mathbf{w}^*, 
     \frac{\eta}{|\mathcal{B}_i|}
    \sum^{|\mathcal{B}_i|}_{j=1} C^{i,j}_t  \nabla F_i(\mathbf{w}_t^i,\xi_j) \rangle}_{\mathcal{C}_2}\\
\end{align*}

It's obvious that $\mathbb{E}[\mathcal{C}_0] = 0$.

By Cauchy-Schwarz inequality and AM-GM inequality, we have
\begin{align*}
    \mathcal{C}_1 &= -2 \sum_{i=1}^N p_i \langle \mathbf{w}_t - \mathbf{w}_{t}^i , \frac{\eta}{|\mathcal{B}_i|}
    \sum^{|\mathcal{B}_i|}_{j=1} C^{i,j}_t  \nabla F_i(\mathbf{w}_t^i,\xi_j) \rangle \\
    &\leq \sum_{i=1}^N p_i \| \mathbf{w}_t - \mathbf{w}_{t}^i \|^2 + \| \sum_{i=1}^N p_i \frac{\eta}{|\mathcal{B}_i|}
    \sum^{|\mathcal{B}_i|}_{j=1} C^{i,j}_t  \nabla F_i(\mathbf{w}_t^i,\xi_j) \|^2\\
    &=  \sum_{i=1}^N p_i \| \mathbf{w}_t - \mathbf{w}_{t}^i \|^2 + \mathcal{B}_1
\end{align*}
So we get
\begin{equation*}
    \mathbb{E}[\mathcal{C}_1] \leq \sum_{i=1}^N p_i \mathbb{E} \| \mathbf{w}_t - \mathbf{w}_{t}^i \|^2 +\eta^2G^2C_t^2
\end{equation*}

According to Assumption \ref{ass:mu_strong_convex}, we know that
\begin{align*}
    -\langle \mathbf{w}_{t}^{i} - \mathbf{w}^* , \nabla F_i(\mathbf{w}_t^i) \rangle \leq - \left( F_i(\mathbf{w}_{t}^{i}) - F_i(\mathbf{w}^*)  \right) - \frac{\mu}{2} \| \mathbf{w}_{t}^{i} - \mathbf{w}^* \|^2
\end{align*}

So we get
\begin{align*}
    \mathcal{C}_2 &\leq 2\sum_{i=1}^N p_i \frac{\eta}{|\mathcal{B}_i|}\sum_{j=1}^{|\mathcal{B}_i|} C_t^{i,j} \cdot 
    \left[ -\left( F_i(\mathbf{w}_t^i,\xi_j) - F_i(\mathbf{w}^*) \right) - \frac{\mu}{2}\| \mathbf{w}_t^i - \mathbf{w}^* \|^2 \right] \\
    &\leq -2\eta C_t \sum_{i=1}^N p_i \left( F_i(\mathbf{w}_t^i) - F_i(\mathbf{w}^*) \right) - \mu\eta C_t \sum_{i=1}^N p_i \| \mathbf{w}_t^i - \mathbf{w}^* \|^2\\
    &= -2\eta C_t \sum_{i=1}^N p_i \left( F_i(\mathbf{w}_t^i) - F^*+F^* - F_i(\mathbf{w}^*) \right)
    - \mu\eta C_t \sum_{i=1}^N p_i \| \mathbf{w}_t^i - \mathbf{w}^* \|^2\\
    &= -2\eta C_t \underbrace{\sum_{i=1}^N p_i \left( F_i(\mathbf{w}_t^i) - F^* \right)}_{\mathcal{D}_1} - 2\eta C_t \Gamma - \mu\eta C_t \| \mathbf{w}_t - \mathbf{w}^* \|^2,
\end{align*}
where $\Gamma= \sum_{i=1}^N p_i (F^*-F_i^*)=F^*-\sum_{i=1}^N p_i F_i^*$.

Next, we proceed to handle $\mathcal{D}_1$.
\begin{align*}
    \mathcal{D}_1 &= \sum_{i=1}^N p_i \left( F_i(\mathbf{w}_t^i) - F_i(\mathbf{w}_t) \right) + \sum_{i=1}^N p_i \left( F_i(\mathbf{w}_t) - F^* \right) \\
    & \geq \sum_{i=1}^N p_i \langle \nabla F_i(\mathbf{w}_t),\mathbf{w}_t^i - \mathbf{w}_t \rangle
    + (F(\mathbf{w}_t) - F^* )  \\
    &\text{(from the Assumption \ref{ass:mu_strong_convex})}\\
    &\geq -\frac{1}{2} \sum_{i=1}^N p_i 
    \left[ \eta \| \nabla F_i(\mathbf{w}_t)\|^2 + \frac{1}{\eta} \| \mathbf{w}_t^i - \mathbf{w}_t \|^2 \right]
    + (F(\mathbf{w}_t) - F^* ) \\
    &\text{(from the AM-GM inequality)}\\
    &\geq - \sum_{i=1}^N p_i \left[ \eta L(F_i(\mathbf{w}_t)-F_i^*) + \frac{1}{2\eta} \| \mathbf{w}_t^i - \mathbf{w}_t \|^2 \right]
    + (F(\mathbf{w}_t) - F^* ) \\
    &\text{(from the L-smooth inference)}\\
    &\geq -(\eta L+1)\Gamma - \frac{1}{2\eta}\sum_{i=1}^N p_i \| \mathbf{w}_t^i - \mathbf{w}_t \|^2,
\end{align*}
where L-smooth inference as show:
\begin{align}
    \|\nabla F_i(\mathbf{w}_t^i)\|^2 \leq 2L\left( F_i(\mathbf{w}_t^i) - F_i^* \right).
\end{align}

Thus, we get
\begin{align*}
    \mathcal{C}_2 \leq 2\eta^2 C_t L \Gamma +
    C_t \sum_{i=1}^N p_i \| \mathbf{w}_t^i - \mathbf{w}_t \|^2
    - \mu\eta C_t \| \mathbf{w}_t - \mathbf{w}^* \|^2
\end{align*}

To sum up,
\begin{align*}
    \mathbb{E}[\mathcal{A}_1] &= \mathbb{E} \sum_{i=1}^N p_i 
    \| \mathbf{w}_t - \mathbf{w}_t^i \|^2
    + \eta^2 G^2 C_t^2 + 2\eta^2 C_t L \Gamma \\
    &\quad + C_t \mathbb{E} \left[\sum_{i=1}^N p_i \| \mathbf{w}_t^i - \mathbf{w}_t \|^2\right]
    - \mu \eta C_t \mathbb{E} \| \mathbf{w}_t - \mathbf{w}^* \|^2 \\
    &= (1+C_t)\mathbb{E} \left[\sum_{i=1}^N p_i 
    \| \mathbf{w}_t - \mathbf{w}_t^i \|^2\right] - \mu\eta C_t \mathbb{E} \| \mathbf{w}_t - \mathbf{w}^* \|^2 \\
    &\quad + \eta^2 \left( G^2 C_t^2 + 2L\Gamma C_t \right),
\end{align*}
and
\begin{align*}
    &\quad \mathbb{E} \left[\sum_{i=1}^N p_i 
    \| \mathbf{w}_t - \mathbf{w}_t^i \|^2  \right]\\
    &= \mathbb{E} \left[\sum_{i=1}^N p_i 
    \| (\mathbf{w}_t - \mathbf{v}_{t+1}^i) - (\mathbf{w}_t^i - \mathbf{v}_{t+1}^i) \|^2  \right] \\
    &\leq \mathbb{E} \left[\sum_{i=1}^N p_i 
    \| \mathbf{w}_t^i - \mathbf{v}_{t+1}^i \|^2  \right]\\
    &\leq \mathbb{E} \left\|\sum_{i=1}^{N}p_i \frac{\eta}{|\mathcal{B}_i|}[
    \sum^{|\mathcal{B}_i|}_{j=1} C^{i,j}_t \cdot \nabla F_i(\mathbf{w}_t^i,\xi_j)
    + \sigma C \cdot \mathcal{N}(0,\mathbf{I})]\right\|^2 \\
    &= \mathbb{E}\left[\mathcal{A}_2  \right] = \eta^2C_t^2G^2+\frac{\eta^2\sigma^2C^2d}{\hat{B}^2}.
\end{align*}

So we get,
\begin{align*}
    \mathbb{E}\left[\mathcal{A}_1  \right] &= (1-\mu\eta C_t) \mathbb{E}\| \mathbf{w}_t - \mathbf{w}^* \|^2
    + (1+C_t)\eta^2C_t^2G^2 
    + \eta^2\left(
    2G^2C_t^2 + 2L\Gamma C_t + 2\frac{\sigma^2 C^2 d}{\hat{B}^2}
    \right)
\end{align*}

All in all, we get
\begin{align*}
    \mathbb{E}\| \mathbf{v}_{t+1} - \mathbf{w}^* \|^2
    \leq (1-\mu\eta C_t) \mathbb{E}\| \mathbf{w}_{t} - \mathbf{w}^* \|^2 +\eta^2 A,
\end{align*}
where
\begin{align*}
    A=G^2 C_t^3 + 3 G^2 C_t^2 + 2L\Gamma C_t + \frac{2\sigma^2 C^2 d}{\hat{B}^2}.
\end{align*}

\section{Related Work}\label{sec:related_work}

While various fairness concepts exist in machine learning~\cite{cong2020game,lyu2020collaborative}, recent studies in DPFL have predominantly focused on \textit{group fairness} to mitigate biases related to sensitive attributes like gender and race~\cite{galvez2021enforcing,padala2021federated,gu2022privacy}. Notable works have integrated DP with fairness constraints (e.g., equal opportunity or decision boundary fairness) to protect attribute privacy while reducing discrimination~\cite{jagielski2019differentially,tran2021differentially,ding2020differentially}.

In contrast, our research shifts focus to \textit{balanced performance fairness}, a relatively unexplored area within DPFL. Unlike group fairness which targets demographic parity, balanced performance fairness—introduced by Li et al.~\cite{TianLi2020qFFL} and further explored in~\cite{zhao2022dynamic,li2021ditto}—prioritizes uniform model performance across clients, making it particularly suitable for federated settings involving diverse institutions. While parameter tuning (e.g., clipping thresholds) can improve group fairness~\cite{gu2022privacy}, achieving balanced performance fairness necessitates fundamentally adjusting the direction of model updates. Balancing these directional adjustments with the noise and clipping inherent in DP presents a unique challenge. To the best of our knowledge, this is the first study to address balanced performance fairness within the context of DPFL.

\end{document}